# Theory of high-energy emission
# from the pulsar/Be-star system PSR 1259−63 I:
# radiation mechanisms and interaction geometry

Marco Tavani

*Columbia Astrophysics Laboratory*
*Columbia University, New York, NY 10027*

and

Jonathan Arons

*Department of Astronomy and Department of Physics*
*University of California at Berkeley, Berkeley, CA 94720*
and
*Institute of Geophysics and Planetary Physics*
*Lawrence Livermore National Laboratory, Livermore, CA 94550*

## Abstract

We study the physical processes in the system containing the 47 ms radio pulsar PSR B1259–63 orbiting around a Be star in a highly eccentric orbit. This system is the only known binary where a radio pulsar is observed to interact with gaseous material from a Be star. A rapidly rotating radio pulsar such as PSR B1259–63 is expected to produce a wind of electromagnetic emission and relativistic particles, and this binary is an ideal astrophysical laboratory to study the mass outflow/pulsar interaction in a highly time variable environment. Motivated by the results of a recent multiwavelength campaign during the January 1994 periastron passage of PSR B1259-63, we discuss several issues regarding the mechanism of high-energy emission. Unpulsed power law emission from the PSR B1259–63 system was detected near periastron in the energy range 1-200 keV. The observed X-ray/soft $\gamma$-ray emission is characterized by moderate luminosity, small and constant column density, lack of detectable pulsations and peculiar spectral and intensity variability. In principle, high energy (X-ray and gamma-ray) emission from the system can be produced by different mechanisms including: (1) mass accretion onto the surface of the neutron star, (2) 'propeller'-like magnetospheric interaction at a small pulsar distance, and (3) shock-powered emission in a pulsar wind termination shock at large distance from the pulsar. We carry out a series of calculations aimed at modelling the high-energy data of the PSR B1259–63 system throughout its orbit and especially near periastron. We find that the observed high energy emission from the PSR B1259–63 system is not compatible with accretion or propeller-powered emission. This conclusion is supported by a model based on standard properties of Be stars and for plausible assumptions about the pulsar/outflow interaction geometry.

We find that shock-powered high-energy emission produced by the pulsar/outflow interaction is consistent with all the characteristics of the high energy emission of the PSR B1259–63 system. This opens the possibility of obtaining for the first time constraints on the physical properties of the PSR B1259–63 pulsar wind and its interaction properties in a strongly time variable nebular environment. By studying the time evolution of the pulsar cavity we can constrain the magnitude and geometry of the mass outflow as the PSR B1259–63 orbits around its Be star companion. The pulsar/outflow interaction is most likely mediated by a collisionless shock at the internal boundary of the pulsar cavity. The system shows all the characteristics of a *binary plerion* being *diffuse* and *compact* near apastron and periastron, respectively. The PSR B1259–63 system is subject to different radiative regimes depending on whether synchrotron or inverse Compton (IC) cooling dominates the radiation of electron/positron pairs ($e^{\pm}$-pairs) advected away from the inner boundary of the pulsar cavity. The highly non-thermal nature of the observed X-ray/soft $\gamma$-ray emission from the PSR B1259–63 system near periastron establishes the existence of an efficient particle acceleration mechanism within a timescale shown to be less than $\sim 10^2 - 10^3$ s. A synchrotron/IC model of emission of $e^{\pm}$-pairs accelerated at the inner shock front of the pulsar cavity and adiabatically expanding in the MHD flow provides an excellent explanation of the observed time variable X-ray flux and spectrum from the PSR B1259–63 system. We find that the best model for the PSR B1259–63 system is consistent with the pulsar orbital plane being misaligned with the plane of a thick equatorial Be star outflow. The angular width of the equatorially enhanced Be star outflow is constrained to be $\sim 50°$ at the pulsar distance, and the misalignment angle is $\gtrsim 25°$. We calculate the intensity and spectrum of the high-energy emission for the whole PSR B1259–63 orbit and predict the characteristics of the emission near the apastron region based on the periastron results. The mass loss rate is deduced to be approximately constant in time during a $\sim 2$ year period. Our results for the Be star outflow of the PSR B1259–63 system are consistent with models of the radio eclipses near periastron.

The consequences of our analysis have general validity. Our study of the PSR B1259–63 system shows that X-ray emission can be caused by a mechanism alternative to accretion in a system containing an energetic pulsar interacting with nebular material. This fact can have far-reaching consequences for the interpretation of galactic astrophysical systems showing non-thermal X-ray and $\gamma$-ray emission. We show that a binary system such as PSR B1259–63 offers a novel way to study the acceleration process of relativistic plasmas subject to strongly time variable radiative environments.

**Subject headings:** pulsars: PSR B1259-63, Be stars, relativistic winds

## 1. Introduction

The PSR B1259–63 system contains a rapidly rotating radio pulsar with spin period $P = 47.76$ ms and spindown luminosity $\dot{E}_R \simeq 8 \times 10^{35}$ erg s$^{-1}$ orbiting around a massive Be star companion (Johnston *et al.* 1992, 1994, 1996; Manchester *et al.* 1995, hereafter M95). The pulsar moves in a highly elliptical orbit with period $\sim 3.4$ yrs. The system's distance from the Earth is $d \simeq 2$ kpc (Johnston *et al.* 1994): note, however, that $d$ is uncertain by a factor of $\sim 2$ based on radio measurements (e.g., Taylor, Manchester & Lyne 1993). Table 1 summarizes the main pulsar and orbital characteristics of the only known binary where the interaction of a pulsar with a mass outflow from a high mass star can be studied in detail. The PSR B1259–63 system turns out to be an important astrophysical laboratory for the study of pulsars interacting with gaseous environments. Be stars are characterized by large mass outflows mainly concentrated in the equatorial plane of the massive star (e.g., Waters *et al.* 1988). Because of the highly elliptical orbit, PSR B1259–63 is expected to interact with the Be star outflow especially near periastron. The nature of the pulsar/outflow interaction can be studied with the help of the high-energy emission (X-rays and $\gamma$-rays) radiated by shock disruption of the Be star outflow (Tavani, Arons & Kaspi 1994, hereafter TAK94). Previous studies of this binary addressed several theoretical issues concerning the PSR B1259–63 system such as its evolution (King 1993; Lipunov *et al.* 1994) and the pulsar/outflow interaction (Kochanek 1993; TAK94; Campana *et al.* 1995; Ghosh 1995; Melatos, Johnston & Melrose 1995).

The main aim of this paper is to provide a detailed model for the high-energy emission (X-ray and soft $\gamma$-ray) originating from this interaction in the PSR B1259–63 system. We will base our analysis on the wealth of multiwavelength information obtained in the X-ray and $\gamma$-ray bands by ROSAT, ASCA and GRO near the apastron and periastron passages of PSR B1259–63 (Cominsky *et al.* 1994, hereafter CRJ94; Kaspi *et al.* 1995, hereafter KTN95; Grove *et al.* 1995, hereafter G95; Tavani *et al.* 1996a, hereafter T96; Hirayama *et al.* 1996, hereafter H96). In particular, the series of repeated ASCA and extended GRO observations of the PSR B1259–63 system near periastron provides invaluable information allowing a detailed theoretical analysis of the relevant radiation mechanisms. ASCA observed and detected the PSR B1259–63 system four times near periastron, including observations before, during, and after periastron (KTN95) during which the radio pulsar PSR B1259–63 was eclipsed by nebular circumbinary material and not detectable. A crucial fourth ASCA observation was carried out in late February 1994, i.e., about $\sim 50$ days after the periastron passage of January 9, 1994 (Hirayama *et al.* 1996). At that time, the radio pulsar was again detectable (M95, Johnston *et al.* 1996. hereafter J96). The *Compton* GRO uninterrupted observation of the PSR B1259–63 system lasted for about three weeks (January 3-23, 1994). The length of the GRO observation was crucial for the OSSE instrument in detecting a signal at the level of a few mCrab in the hard X-ray energy range (G95). The absence of detectable pulsed radio emission near periastron clearly indicates the presence of circumbinary scattering/absorbing material.

High-energy emission from the PSR B1259–63 system can be produced by several different mechanisms: Be star coronal emission, pulsar magnetospheric emission, accretion onto the surface of the neutron star, interaction with the pulsar magnetosphere in a propeller regime (Illarionov & Sunyaev 1975, King & Cominsky 1994, Campana *et al.* 1995, Ghosh 1995), bremsstrahlung emission of heated material in the Be star outflow (*e.g.*, Kochanek 1993), non-thermal emission at the pulsar wind termination shock (TAK94). A primary aim of this paper is to check the plausibility of these different emission processes based on the available high-energy data. As we will see, only magnetospheric processes or shock-driven emission survive as viable mechanisms after a first simple analysis. Distinguishing between magnetospheric and shock driven processes requires a more detailed study of the flow. In a separate paper (Tavani *et al.* 1996a) we show that nonthermal emission from the magnetosphere is also excluded. The second part of the paper is devoted to a systematic study of the non-thermal emission from the region just downstream from the relativistic shock terminating the pulsar's wind, the remaining viable candidate.

The PSR B1259–63 system is a complex physical system influenced by the combined effect of: (A) the Be star outflow, (B) the hydrodynamic and geometric properties of the pulsar/outflow interaction, (C) the characteristics of the pulsar wind, and (D) the non-thermal particle acceleration mechanism at the pulsar wind termination shock. We show in this paper that as a consequence of the large wealth of information obtained on the PSR B1259–63 system (especially near periastron), these different components may be successfully modelled. In this paper, we restrict the discussion primarily to aspects (A) and (B), postponing a detailed discussion of points (C) and (D) to a companion paper (Tavani & Arons, 1996, hereafter Paper II).

We proceed as follows. Sect. 2 contains a brief summary of the relevant multiwavelength data on the PSR B1259–63 system with a discussion of possible emission mechanisms at work. Sect. 3 summarizes the main properties of Be star mass outflow that are relevant for our tasks. Sect. 4 contains a brief summary of the properties of relativistic pulsar winds relevant for PSR B1259-63. We will be equipped to confront the first relevant calculation of our paper given in Sect. 5 on the properties of the pulsar cavity as a function of orbital phase for different assumptions on the Be star outflow. A discussion of the different high-energy emission mechanisms is then presented in Sect. 6. Crucial use of ASCA and GRO information obtained near the periastron passage is made to discriminate between accretion, magnetospheric and shock-powered models.

The second part of the paper is devoted to a general discussion of different non-thermal radiative regimes of the pulsar cavity as a function of orbital phase and pulsar/outflow interaction geometry. Sect. 7 presents a general discussion of the relevant processes and timescales and Sect. 8 gives the main formulae for the non-thermal emission luminosity for different radiative regimes. Sect. 9 presents the theory of non-thermal radiation in the presence of strong synchrotron and inverse Compton cooling of accelerated $e^{\pm}$-pairs of the pulsar wind. The calculation is carried out in general terms, with no specific reference to a particle acceleration mechanism except for the assumption (justified by the highly non-thermal observed high-energy spectrum) of efficient and fast non-thermal acceleration at or near the pulsar wind termination shock. Sect. 10 gives a comparison of theoretically calculated and observed X-ray emission and spectral properties of the PSR B1259–63 system. We show that the geometry of the PSR B1259–63 orbit vs. the Be star equatorial outflow (coplanar or misaligned model) can be sensibly constrained by the time dependence of the intensity and spectral variation of the X-ray emission. Sect. 11 contains a discussion and our conclusions on the general applicability of our results for the interpretation of the nature of galactic high-energy sources.

In this paper we make use of the results obtained in Paper II concerning the determination of the relativistic pulsar wind parameters for the PSR B1259–63 system. We find in Paper II that the pre-shock value of the Lorentz factor of pulsar wind particles of PSR B1259–63 is $\gamma_1 \gtrsim 10^6$, and that the pulsar wind magnetization is $\sigma \sim 0.1$-$0.01$. Even though our primary objective is the modelling of the PSR B1259–63 system, our treatment of the pulsar/outflow interaction and of radiative processes is given in a way that it can be easily generalized to other systems.

## 2. Summary of observations of the PSR B1259–63 system

The radio pulsar PSR B1259–63 of spin period $P = 47.76$ ms was discovered in a high frequency survey of the Southern Galactic plane (Johnston *et al.* 1992). Pulsar timing confirmed by later observations revealed that the pulsar is in a high-eccentricity, long-orbital period binary system with a massive companion (J96; M95). The updated pulsar and orbital parameters are given in Table 1. Its companion star is the Be-star SS 2883, a 10th magnitude Be star of luminosity $L_* = 5.8 \times 10^4 \, L_\odot$, estimated radius $R_* \sim (6-10) \, R_\odot$ and distance from Earth between 1.5 and 3 kpc (Johnston *et al.*, 1994). This luminosity and radius of the Be star companion correspond to an effective temperature of $T_R \simeq 27,000$ K at the star's surface. Only a lower bound for the mass $M$ of SS 2883 is known from pulsar observations. For a plausible mass $M = 10 \, M_\odot$ (that we assume in our calculations), the pulsar orbit inclination angle with respect to the plane of the sky is $\sim 35°$ (J96).

Extensive radio observations of PSR B1259–63 carried out in 1993 and 1994 (J96) showed a first sign of pulsar/outflow interaction in the PSR B1259–63 system in October 1993 (approximately at $\mathcal{T} - 94$, where $\mathcal{T} =$ January 9.2, 1994, TJD 49361.7 is the periastron date). Substantial radio pulse depolarization was observed, with occasional eclipsing behavior at 1.5 GHz at the end of November 1993. The pulsar was not detected in 1.5 GHz data on Dec. 20, 1993 ($\mathcal{T} - 20$) and reappeared on Feb. 4, 1994 ($\mathcal{T} + 24$). During the 44-day eclipse PSR B1259–63 was not detected in extensive observations at 1.5 and 8.4 GHz (J96). The dispersion measure showed substantial increase in the pre-eclipse data, reaching a value of $\Delta DM \sim 11 \, \text{cm}^{-3}$ pc above normal, corresponding to an average free electron column density of $N_e \sim 3.3 \cdot 10^{19} \, \text{cm}^{-2}$. By mid-April 1994, both the linear polarization and the rotation measure were back to pre-eclipse values. A change of pulsar/orbital parameters as determined by radio data was observed in coincidence with the last two periastron passages of PSR B1259–63 (M95). The parameter change manifests itself as an apparent spindown of the pulsar spin period with $\Delta P/P \sim +10^{-9}$ or as a change of orbital parameters of order $10^{-5}$ (note that pulsar and orbital parameters are strongly covariant for the PSR B1259–63 system, see M95). An interpretation in terms of spindown torque by a propeller effect (M95) requires very special conditions for the gaseous matter to be accumulated near the corotation radius of PSR B1259–63 (a very small coupling coefficient between the pulsar wind and the Be star outflow $\xi_p \lesssim 10^{-4}$, and accumulated mass $m_{acc} \sim 2 \cdot 10^{19}$ g over a period of $\sim 20$ days for a resulting radiated X-ray luminosity of $\sim 10^{31}$ erg s$^{-1}$). We will discuss the plausibility of these assumptions in the next sections.

Optical observations of the PSR B1259–63 system near periastron show little perturbation of the broad H$\alpha$ emission profile believed to have its origin in the Be star equatorial outflow (Manchester 1994). This can be interpreted as an indication of a minor perturbation of the Be star outflow disk at periastron.

Since its discovery in 1992, the PSR B1259–63 system was observed several times by X-ray instruments. Three ROSAT observations were carried out in 1992-1993 when PSR B1259–63 was near its apastron (Cominksy, Roberts & Johnston 1994; hereafter CRJ94; Greiner, Tavani & Belloni 1995). Four ASCA observations of PSR B1259–63 covered the time during the 1994 periastron passage (KTN95; T96). An extended three-week observation near periastron was carried out by the *Compton* Gamma-Ray Observatory (GRO) (T96; G95). Fig. 1 shows graphically the timing of X-ray observations of PSR B1259–63 along its orbit; Fig. 2 summarizes the ROSAT and ASCA detected fluxes. The PSR B1259–63 system was always detected by ROSAT and ASCA, and a previously obtained upper limit by GINGA is consistent with a monotonic decrease of the X-ray emission as the pulsar moves from periastron to apastron. X-ray ASCA observations of the PSR B1259–63 system near periastron (KTN95; H96) can be summarized as follows:

(1) the source is always detected with moderate luminosity (in the energy range 1-10 keV) $L_x \sim 10^{34}$ erg s$^{-1}$ corresponding to $\sim 1\%$ of the pulsar spindown luminosity. The X-ray intensity within single $\sim 1$ day observations is constant. The X-ray flux is not constant among different observations $\sim 2-4$ weeks apart from each other. The flux appears to be varying by a factor of $\sim 2$ with a minimum at periastron and two local maxima occurring about 15 days before and after the periastron passage (see Fig. 2);

(2) the X-ray spectrum is consistent with a power-law of photon index $\alpha_X \sim 1.6-1.9$; no X-ray lines are detected. The photon index is not constant across the periastron observations, but it varies as a function of time. The spectrum is substantially steeper at periastron ($\alpha_X \sim 1.9$) and harder before and after periastron (see Fig. 3);

(3) the system is characterized by a constant small photoelectric column density $N_H \sim 6 \cdot 10^{21}$ cm$^{-2}$, which is approximately equal to the value deduced for the 1992-1993 apastron observations (CRJ94) (see Fig. 3);

(4) no significant X-ray pulsations with the pulsar spin period were detected for all ASCA observations. The 90% confidence upper limit for the amplitude of a possible sinusoidal modulation is $\sim 7\%$.

Figs. 3 and 2 show the behavior of the column density and spectral hardness of the X-ray emission as well as the luminosity evolution along the orbit, respectively. It is clear from Fig. 2 that the X-ray luminosity increases by one order of magnitude as the pulsar approaches periastron with a non-monotonic behavior as a function of orbital phase. There is also a clear tendency for the emission to decrease as the pulsar recedes from periastron.

OSSE is the only GRO instrument which detected high-energy emission from the PSR B1259–63 system during an extended observation near periastron (January 3-23, 1994). The source has been detected in the 30-200 keV range at the level of a few mCrab $[(2.8 \pm 0.6) \cdot 10^{-5}\, \mathrm{ph\, cm^{-2}\, s^{-1}\, MeV^{-1}}$ at 100 keV], with a spectrum consistent with a power law of photon index $\alpha_\gamma \sim 1.8$ (G95). For a 2 kpc distance (J94), the inferred luminosity is $L_{x,hard} \simeq 3 \cdot 10^{34}\, \mathrm{erg\, s^{-1}}$, i.e., $\sim 4\%$ of the total pulsar spindown luminosity. No time variability within the three-week observing period could be determined due to the weakness of the source. It is important to note that the OSSE spectrum agrees with the time-averaged ASCA spectrum extrapolated in the hard X-ray range (G95, T96). A contribution from the galactic diffuse background is estimated to be $\sim 10$ times smaller than the detected emission. The flux detected by OSSE is unpulsed, and the upper limits for pulsed emission of phase width 0.15 and 0.5 are $3 \cdot 10^{-4}$ and $7 \cdot 10^{-4}\, \mathrm{ph\, cm^{-2}\, s^{-1}\, MeV^{-1}}$, respectively. An OSSE pointing at the PSR B1259–63 system carried out in September 1991 resulted in an upper limit for pulsed emission of $6 \cdot 10^{-3}\, \mathrm{ph\, cm^{-2}\, s^{-1}\, MeV^{-1}}$ (Ray *et al.* 1993).

No other GRO instrument detected the PSR B1259–63 system despite extensive searches with BATSE, COMPTEL and EGRET data of unpulsed and pulsed emission (T96). The EGRET upper limit to unpulsed emission is particularly relevant (the 95% confidence limit in the 100-300 MeV band is $4.8 \cdot 10^{-11}\, \mathrm{erg\, cm^{-2}\, s^{-1}}$) because it clearly implies a spectral cutoff in the MeV range (T96). Fig. 4 shows a collection of the ASCA and GRO spectral results obtained near periastron.

## 2.1. Possible Origins of X-ray Emission in the PSR B1259–63 system

High-energy emission from the PSR B1259–63 system can be produced by several different mechanisms:

**(1)** *Accretion* onto the surface of the neutron star can occur with consequent bright X-ray/hard X-ray emission for large mass outflow rates and high density of material near periastron/footnote A remarkable example of a related source is given by the 69 ms X-ray pulsar A0538-66 in the Large Magellanic Cloud which occasionally accretes producing an X-ray luminosity near the Eddington limit $L_E \sim 10^{38}\, \mathrm{erg\, s^{-1}}$ (Skinner *et al.* 1982)..

**(2)** *Propeller effect* emission, caused by gaseous material being temporarily trapped between the light cylinder and corotation radius of the pulsar (e.g., Illarionov & Sunyaev 1975; Lipunov *et al.* 1994; Campana *et al.* 1994; Ghosh 1995);

(3) *Pulsar/outflow interaction*, caused by the shock of a relativistic pulsar wind in the mass outflow from the companion star can produce unpulsed high-energy emission (TAK94).

(4) *Pulsed magnetospheric emission*. The pulsar age of PSR B1259–63 is relatively large ($\tau_p \equiv P/2\dot{P} \sim 3.3 \cdot 10^5$ yrs) and the inferred dipolar magnetic field is sufficiently small ($B \sim 3.3 \cdot 10^{11}$ G) to allow an interesting comparison of the detection (or lack thereof) of pulsed high-energy emission from PSR B1259–63 compared to other pulsars of similar age (T96).

(5) *Be star coronal emission*. The observed value of the ratio of X-ray to bolometric luminosities near periastron is $\sim 10^{-4}$, i.e., on the high side of the observed distribution of ratios for OB and OBe stars (e.g., Meurs *et al.* 1992). A typical spectrum of Be star coronal emission is quite soft with a typical coronal temperature of $T \sim (2-9) \cdot 10^6$ K (Cassinelli *et al.* 1994). Both the spectral and time-variability features of the PSR B1259–63 system near periastron are therefore in disagreement with typical Be star coronal emission.

(6) *Bremsstrahlung emission by heated gas of the Be star outflow* (Kochanek 1993). We find that the emission measure of shock-heated gas of the Be star outflow and its expected temperature are both in disagreement with the intensity and spectrum of the hard power-law emission up to $\sim 200$ keV observed from the PSR B1259–63 system near periastron.

In the following we will discuss in detail mechanisms (1-3), leaving aside the issue of pulsed magnetospheric emission from PSR B1259–63 (T96). From ASCA data, the pulsed magnetospheric X-ray luminosity of PSR B1259–63 is constrained to be $\lesssim 10^{33}$ erg s$^{-1}$ (KTN95, H96).

## 3. Properties of Be star outflows

Be stars rotate rapidly, producing highly asymmetric mass loss. Both a dilute 'polar' component and a denser 'equatorial' component of outflow have been invoked to reconcile models for the IR and optical observations of Be stars (Waters *et al.*, 1988; hereafter W88). For the equatorial 'disk-like' component, the density near the star's surface can exceed that near the pole by a factor of $\sim 100$ or more (W88, Bjorkman & Cassinelli 1993). For our purposes, the parametrizations of Be star outer flow employed in the literature will suffice to characterize the properties of the mass and momentum outflow of SS 2883. The high-density equatorial disk wind region has a density profile (derived from the IR excess) $\rho(R)$, of the form

$$\rho(R) = \rho_0 (R/R_\star)^{-n}, \tag{1}$$

where $R_\star$ is the radius of the star (Waters 1986; W88), $R$ the distance from the Be star surface (equal to $R = d - r$, with $r$ the radial distance from the pulsar and $d$ the separation between the two stars), and $\rho_0 = \dot{M}/4\pi f_w R_\star^2 v_0$, with $\dot{M}$ the mass outflow rate, $f_w$ the fraction of $4\pi$ sterradians into which the equatorial wind flows, and $v_0$ the velocity at the stellar surface. Typically, $\rho_0$ is of the order of $10^{-10} - 10^{-13}$ g cm$^{-3}$, the Be star radius is of order $R_\star \sim 10\,R_\odot$, and the 'outflow exponent' is in the range, $2 < n < 4$. In some cases, there is evidence for the presence of disk material out to at least several hundreds of stellar radii. In the case of SS 2883, analysis of the radio eclipse near periastron suggests the presence of disk material up to several tens of stellar radii (Melatos *et al.* 1995).

From the assumed behavior of density of the outflow as a function of distance from the Be star surface, the equation of continuity for a steady wind flowing into an angular sector with solid angle $4\pi f_w$ steradians, $\dot{M} = 4\pi f_w \rho R^2 v_w$, yields the velocity field

$$v_w(R) = v_0 \left(\frac{R}{R_\star}\right)^{n-2},$$   (2)

where $v_0$ is believed to be in the range $10^6$–$10^7$ cm s$^{-1}$ for typical Be star outflows (W88). As a consequence, the dynamic outflow pressure at distance $R$ from the surface of the Be star has a radial dependence of the form

$$P_w(R) = \rho(R)\,v_w(R)^2 = \frac{\dot{M}v_0}{4\pi f_w R_\star^2} \left(\frac{R}{R_\star}\right)^{n-4}$$   (3)

We assume the flow to be supersonic, as is consistent with the observed temperatures of the outflow, which correspond to a sound speed $c_s \sim 10$ km/s $\ll v_w(R)$. Physically meaningful cases correspond to $n < 4$, for which the dynamic pressure decreases monotonically with increasing $R$. In the following, we will parametrize the properties of the Be star mass outflow by specifying four parameters ($\dot{M}$, $v_0$, $n$, $f_w$). However, for supersonic outflows, the role of the Be star wind is to stop and possibly confine the pulsar wind, so that only the dynamic pressure $P_d$ enters into our model construction. From Eq. 3, $P_d$ depends only on the momentum loss rate $\dot{M}v_0$, not on $\dot{M}$ and $v_0$ separately. Therefore, the role of the Be star wind can be fully parameterized by the outflow parameter $\Upsilon$ defined as in TAK94

$$\Upsilon = \dot{M}_{-8}\,v_{0,6}$$   (4)

where $\dot{M} = (10^{-8}\,M_\odot\,\mathrm{yr}^{-1})\,M_{-8}$, and $v_0 = (10^6\,\mathrm{cm\,s}^{-1})\,v_{0,6}$. In the following, we will mostly consider the range $1 \lesssim \Upsilon \lesssim 10^3$ and show that regimes with $\Upsilon \gg 10^3$ are not supported by observations of the PSR B1259–63 system. The separate parameters $n$ and $f_w$ are both constrained by the observations.

### 3.0.1.   *Pulsar orbit inclination with respect to the Be-star equator*

The PSR B1259–63 system presumably originated from a Type II supernova event forming the pulsar. Depending on the asymmetry of the supernova explosion, a the newly formed neutron star can be substantially kicked out of the original orbital plane (e.g., Hills 1983). In principle, the inclination angle between the pulsar orbit and the Be star equatorial plane can be quite large. Several long-period Be star–X-ray binary systems of moderate eccentricities provide evidence for small inclination angles between the orbital plane and the Be stars' rotational equators. They are persistent X-ray sources having luminosities implying that the neutron star is embedded in a high-density region, (assumed to be the Be star equatorial plane) throughout its orbit (Waters *et al.*, 1989). However, the very high-eccentricity PSR B1259–63 system provides an ideal example of a binary where a substantial kick of the neutron star above the original orbital plane might have occurred, which has not yet had time to dissipate. Neutron star 'splashing' into a Be star equatorial disk on its way into and out of periastron in a misaligned geometry could produce a 'double hump' X-ray signal, if the X-rays arise either from accretion or from the interaction of the pulsar-Be star winds. It is therefore necessary to consider two possible models for the PSR B1259–63 system: a *coplanar model* characterized by a small inclination angle between the pulsar orbit and the Be star equatorial outflow, and a *misaligned* model for a substantial inclination angle ( $\gtrsim 10°$ ).

For misaligned pulsar/outflow models we will consider the parametrization with two outflow parameters, $\Upsilon_1$ and $\Upsilon_2$ reflecting the momentum flow in the polar wind and the equatorial outer disk, respectively. The outflow parameter $\Upsilon_1$ will be assumed to be independent of geometry. However, the effective outflow parameter $\Upsilon_2$ strongly depends on the thickness of the Be star equatorial flow. We parametrize the angular dependence of $\Upsilon_2$ along the pulsar orbit as

$$\Upsilon_2 = \Upsilon_2' \left[ \exp[-(\theta - \theta_1)^2/2\sigma_d^2] + \exp[-(\theta - \theta_2)^2/2\sigma_d^2] \right] \tag{5}$$

where $\Upsilon_2'$ is a constant, $\theta$ the orbital phase (defined as $\theta = \phi + \pi$, with $\phi$ the true anomaly), $\sigma_d$ the disk opening angle at the pulsar distance[1], and $\theta_1$ and $\theta_2 = \theta_1 + \pi$ two properly defined phases corresponding to the 'splashing' of the pulsar into the equatorial disk of the Be star companion, the first corresponding to the pulsar's approach to periatron, the second corresponding to its recession from periastron. From Be star observations, we know that the ratio of equatorial to polar momentum flux is in the range $10 \lesssim \Upsilon_2'/\Upsilon_1 \lesssim 10^3$ (W88). Any misaligned model for the PSR B1259–63 system will require the specification of $\Upsilon_1$ and $\Upsilon_2', \theta_1, \sigma_d$.

### 3.0.2. *Time variability of Be-star outflows*

Be stars are known to produce time variable mass loss over a long timescale range (e.g., Coe *et al.* 1993). The most interesting case for our study is provided by A0538-66, a transient source showing 69 ms X-ray pulsations during major accretion episodes as detected in 1980-1982 (Skinner *et al.* 1981). No other major outbursts have been detected from A0538-66 following its discovery. The A0538-66 system might contain a 69 ms pulsar only occasionally able to accrete material from its Be star companion. In the case of the PSR B1259–63 system, we will make the simple assumption that the SS 2883 mass loss does not appreciably vary within one single pulsar orbit. We will show in the following sections that this assumption is consistent with all available observations of the PSR B1259–63 system. However, variations of the SS 2883 mass loss rate might occur at any time in the future. Our formalism can be easily adapted to treat time variable mass loss in pulsar binaries.

---

[1] We assume in the following a constant value of $\sigma_d$ independent of distance from the Be star companion for $R > 10^{12}$ cm.

## 4. Characteristics of Pulsar Winds and Their Termination Shocks

Pulsars lose rotational energy in the form of relativistic MHD winds (e.g., Michel 1969, Arons 1992). The wind carries energy flux in the form of electromagnetic fields and plasma kinetic energy. The energy flux $\mathcal{F}_w$ in the pulsar wind at distance $r \gg R_{lc}$ (where the light cylinder radius of PSR B1259–63 is $R_{lc} = cP/2\pi \simeq 2.28 \cdot 10^8$ cm) is

$$\mathcal{F}_w = \frac{\dot{E}_R}{4\pi f_p r^2} = \frac{c\, B_1^2}{4\,\pi} \left( \frac{1+\sigma}{\sigma} \right) \tag{6}$$

where, using the PSR B1259–63 pulsar parameters of Table 1, $\dot{E}_R = I(2\pi)^2 \dot{P}/P^3 \simeq (8 \times 10^{35} \ \text{erg s}^{-1}) I_{45}$ is the PSR B1259–63 spindown luminosity with $I = 10^{45} I_{45}$ g cm$^2$ the pulsar moment of inertia, and $f_p = \Delta\Omega_p/4\pi$ is the fraction of the sky into which the pulsar wind is emitted. Here $\sigma$ is the pulsar wind magnetization, i.e, the upstream ratio of electromagntic energy density and particle kinetic energy density,

$$\sigma \equiv \frac{B_1^2}{4\pi[N_i m_i + m_\pm(N_+ + N_-)]\gamma_1 c^2} \tag{7}$$

with $\gamma_1$ the Lorentz factor of the MHD particle outflow, and $N_i$ and $N_\pm$ the number densities of ions and electron-positrons with mass $m_i$ and $m_\pm$, respectively. The MHD model of the pulsar wind for the Crab Nebula implies $\sigma \sim 0.005$ (Kennel & Coroniti 1984, hereafter KC84; Emmering and Chevalier 1987). Both the average upstream Lorentz factor $\gamma_1$ and the pulsar wind magnetization $\sigma$ are complex functions of the poorly understood initial acceleration and dynamical evolution of the MHD wind. In the following, we use the results of Paper II concerning the pulsar wind of PSR B1259-63, adopting the set of parameters that best fit the PSR B1259–63 system

$$\gamma_1 = 10^6 \qquad \sigma = 0.02 \ . \tag{8}$$

The pulsar wind's composition ($e^\pm$-pairs, ions) will be addressed in detail in Paper II . However, for completeness, we remark here that if the wind's energy flux is mostly carried by heavy ions, as has been suggested in the case of the Crab pulsar by Hoshino *et al.* (1992, hereafter HAGL92) and by Gallant and Arons (1994, herafter GA94), and if these ions experience close to the maximum magnetospheric potential drop $\Phi \approx (L_p/c)^{1/2} = (1.5 \cdot 10^{15})$ Volts, then $\gamma_1 \approx 10^6$ is the natural Lorentz factor for the wind. This is one of the main arguments in favor of the wind from PSR B1259-63 being ion dominated. Likewise, $\sigma \ll 1$ has been inferred to be a sufficient condition for efficient nothermal acceleration of electrons and positrons in the quasi-transverse relativistic shock waves that terminate pulsar winds (KC84, HAGL92). The theory of pulsar energy loss is not yet capable of predicting $\sigma$ as a function of distance from a pulsar, so the specific value adopted here comes from the analysis described in Paper II .

In principle a pulsar wind may be far from being spherically symmetric. A recent HST/ROSAT study (Hester *et al.* 1995) of the inner regions of the Crab Nebula shows that the cavity blown by the wind in the center of the Nebula fully surrounds the pulsar, with a cavity radius over the pulsar's poles not differing by more than a factor of two from the cavity's equatorial radius, even though the equatorial flow gives rise to the bright X-ray torus while the polar flow shows features reminiscent of "jets". The quasisphericity of the cavity suggests the dynamic pressure of the wind is roughly spherically symmetric, even though latitudinal variations of the emissivity, perhaps traceable to latitudinal variations of wind composition and magnetization, give rise to a much more anisotropic *appearance* of the wind (Arons 1996). The shapes of pulsar bow shock nebulae (Cordes 1996) are also consistent with *approximately* spherical symmetric momentum loss from pulsars. Therefore, as a first approximation, we assume spherical symmetry for the wind from PSR B1259-63. We assume that if the pulsar wind is strong enough to stop accretion from the Be star equatorial disk, it is strong enough to stop infall along all accretion paths.

In principle, a strongly anisotropic pulsar wind pressure might lead to gas penetration towards the pulsar light cylinder as a function of pulsar spin axis inclination angle with respect to the Be star equatorial disk plane. This could result in substantial variations of the detected column density as a function of orbital phase. Also, for quite modest accretion to distances within the light cylinder, one expects the pulsar's collective radio emission to be quenched, or altered to have properties quite different from ordinary radio pulsars. As we will show, there is no evidence for any of these effects in the PSR B1259-63 system, and we conclude that at the distance of $\sim 10^{11}$ cm or more the pulsar wind of PSR B1259-63 is consistent with being spatially isotropic. Thus, while we keep $f_p$, the anisotropy factor for the pulsar wind, in our subsequent formulae, our numerical results all use $f_p = 1$.

Another important issue regarding the modelling of the PSR B1259-63 system concerns the hydrodynamical coupling between the relativistic MHD pulsar wind and its nebular surroundings. Observations and theory of pulsars interacting with interstellar media or binary companions are all consistent with a coupling constant $\xi_p$ of order unity (CK84, Kulkarni & Hester 1988 Cordes, Romani & Lundgren 1993, Bell *et al.* 1995, Brookshaw & Tavani 1995, Cordes 1996) - the momentum of these pulsars' outflows gets deposited in the surroundings as if the interacting media are perfect (magnetized) flows. As shown more extensively in Paper II, the relatively small Larmor radii in the shocked pulsar wind and the outer disk justify the assumption $\xi_p \sim 1$, and previous statements supporting the possibility of very small coupling $\xi_p \ll 1$ (e.g., Kochanek 1993, M95) are not supported by the high-energy data on PSR B1259-63.

## 5. Orbital Evolution of the PSR B1259−63 Pulsar Cavity

A pulsar cavity forms as a result of interaction of the pulsar wind with the mass outflow from the Be star companion of PSR B1259-63. Dynamic pressure balance determines the location of the apex of the pulsar cavity, i.e., of the shock region between the pulsar and Be star outflow:

$$\frac{\dot{E}_R}{4\,\pi\,f_p\,c\,r^2} = \rho(R)v_w(R)^2 \tag{9}$$

where $r$ is the radial distance from the pulsar, and $\rho(R)$ and $v_w(R)$ are the Be star's wind density and velocity, which are functions of $R$, the distance from the center of the Be star (obviously $r$ and $R$ are related by $d = r + R$, with $d$ the orbital distance). Eqs. 1 and 2 describe the radial dependence of the Be star wind density and velocity. Substituting these expressions into Eq. 9, we obtain

$$\frac{\dot{E}_R}{4\,\pi\,f_p\,c\,r^2} = \rho_0 v_0^2 \left(\frac{R}{R_\star}\right)^{n-4} \tag{10}$$

for the distances $R_s$ and $r_s$ from the Be star and the pulsar respectively at which the dynamic pressure between the winds balance, assuming *no* orbital motion of the neutron star with respect to the Be star. We will therefore assume

$$r_s(\phi) + R_s(\phi) = d(\phi), \tag{11}$$

where $\phi$ is the true anomaly. Note that Eq. 11 neglects the difference between the shock radius along the line connecting the pulsar and the Be star center of mass and the shock radius calculated below along the pulsar direction of orbital motion. As long as $r_s \ll d$, this approximation is valid. Eq. 10 can be rewritten as (TAK94)

$$\left[\frac{r_s}{d(\phi)}\right]^2 \left[1 - \frac{r_s}{d(\phi)}\right]^{(n-4)} = \frac{\dot{E}_R}{cf_p} \frac{f_w}{\dot{M}v_0} \left[\frac{R_\star}{d(\phi)}\right]^{n-2} \simeq \frac{40f_w}{f_p \Upsilon(\phi)} \left[\frac{R_\star}{d(\phi)}\right]^{n-2}. \tag{12}$$

We have assumed $R_\star \ll d$, with $d$ the pulsar-Be star separation. In general, $r_s$ is a function of the true anomaly $\phi$, $n$, and $\Upsilon$. In Eq. 12 the ratio

$$\lambda \equiv \frac{\dot{E}_R/cf_p}{\dot{M}v_0/f_w} \simeq \frac{40f_w}{f_p \Upsilon(\phi)}$$

is the ratio of the momentum flux in the pulsar wind to the momentum flux in the Be star wind. Notice that Eq. 12 is a general relation that needs to be satisfied by any interacting flows satisfying generic assumptions on symmetry. For the special case $n = 2$, the solution of Eq. 12 is trivial, $r_s/d = (1 + \lambda^{-1/2})^{-1}$. For a generic index $n$, Eq. 12 needs to be solved numerically.

However, because of the pulsar orbital motion, the outflow velocity relative to the pulsar is not $v_w(R)$. The pulsar orbital motion introduces a 'distortion' of the solution for $r_s$ as the pulsar approaches (recedes) periastron (apastron). The radial ($v_{p,r}$) and tangential ($v_{p,\phi}$) components of the pulsar orbital velocity are

$$v_{p,r} = v_K \frac{e \sin(\phi)}{\sqrt{1-e^2}} \equiv v_K f_r(\phi), \tag{13}$$

$$v_{p,\phi} = v_K \frac{1 + e \cos\phi}{\sqrt{1-e^2}} \equiv v_K f_\phi(\phi), \tag{14}$$

with $\phi$ the orbital phase defined to be zero at periastron[2], and $v_K = \sqrt{G(m_1 + m_2)/a}$ the Keplerian velocity for a binary of total mass $m_1 + m_2$ and semi-major axis $a$. By assuming a Be star outflow velocity of radial component $v_{w,r}$ and tangential component $v_{w,\phi}$, the pressure balance equation becomes

$$\frac{\dot{E}_R}{c(\dot{M}v_o)} \frac{f_w}{f_p} \frac{1}{(d-R_s)^2} = \rho(R_s) \left\{ [v_{w,r}(R_s) - v_{p,r}(\phi)]^2 + [v_{w,\phi} - v_{p,\phi}(\phi)]^2 \right\} \tag{15}$$

The solution of Eq. 15 for the shock radius $R_s$ (or equivalently, for $r_s = d - R_s$) will depend on the orbital phase. By assuming that the Be star outflow velocity is mostly radial, we obtain

$$\frac{1}{(d/R_\star - R/R_\star)^2} = \mathcal{W}(R/R_\star, n, v_K/v_o, \phi) \tag{16}$$

with $\mathcal{W}$ the function

$$\mathcal{W}(R/R_\star, n, v_K/v_o, \phi) \equiv \frac{40}{\Upsilon(\phi)} \frac{f_w}{f_p} \left(\frac{R}{R_\star}\right)^{-n} \left\{ \left[ \left(\frac{R}{R_\star}\right)^{n-\tilde{k}} - \frac{v_K}{v_o} f_r(\phi) \right]^2 + \left[ \frac{v_K}{v_o} f_\phi(\phi) \right]^2 \right\}. \tag{17}$$

We set $\tilde{k} = 2$ for spherical outflow geometry and $\tilde{k} = 1$ for purely disk-like outflow geometry of the Be star wind. Eq. 16 gives the solution of the 'apex' of the pulsar cavity radius aligned along the direction of motion of the pulsar. The lateral dimensions of the pulsar cavity (that can be open or 'closed' depending on the hydrodynamics) can differ by a factor $\sim 2$ compared to the value of Eq. 16 for a constant surrounding pressure (e.g., Baranov *et al.*, 1971). The pulsar cavity size along the line of sight might substantially differ from the calculated $r_s$ depending on the relative inclination angle of the pulsar and Be star disk planes. In the following, we include the effect of the pulsar orbital motion and assume for the typical size of the pulsar cavity the solution $R = R_s$ of Eq. 16. It is understood that the size of the pulsar cavity perpendicular to the pulsar orbital direction may substantially differ from the solution of Eq. 16. In particular, the pulsar wind may be powerful enough to 'break open' the surrounding nebular disk near the interaction ('splashing') points with the Be star outflow.

---

[2]Note that in the figures we use the angle $\theta = \phi + \pi$.

Figs. 5-8 give a few examples of solutions of the Eq. 16. The distance of the shock region from the pulsar, for certain combinations of $B$, $\Upsilon$, and $n$ may approach the quantity $d(\theta) - R_\star$, i.e., the shock might be close to the surface of the Be star. The Be star mass outflow may, in this case, be disrupted by the presence of the pulsar[3].

The magnetic field in the pulsar wind at the shock distance corresponding to the apex of the pulsar cavity can be written as

$$
\begin{aligned}
B_1(r_s) &= \left(\frac{\sigma}{1+\sigma}\right)^{1/2} \left(\frac{\dot{E}_R}{c\, f_p}\right)^{1/2} \frac{1}{r_s} \\
&\approx 0.6 \left[\frac{\sigma}{0.02} \frac{n_{10} v_7^2}{(1+\sigma)}\right]^{1/2} \quad \text{G}.
\end{aligned}
\tag{18}
$$

where we used a typical value the density and gas velocity at the shock $n_{10} = n(R_s)/(10^{10}\,\text{cm}^{-3})$ and $v_7 = v_w(R_s)/(10^7\,\text{cm}\,\text{s}^{-1})$. Note that the magnetic field strength at $r_s$ for a given value of $\sigma$ depends on the value of the dynamic pressure of the Be star's wind at $r_s$ as measured in the frame of an observer moving with the pulsar's orbital velocity. This pressure is independent of the pulsar spindown power, in the approximation that photons and particles, either from the pulsar directly or from the shocked pulsar wind, do not affect the properties of the Be star wind.

The shock in the pulsar wind compresses the magnetic field, which achieves a value $B_2(r_s)$ within the shock thickness (here conservatively assumed to be of order of the ions' Larmor radius $R_L$). The asymptotic value of the downstream magnetic field can be obtained from the relativistic MHD shock conditions (KC84)

$$
B_2(r_s) = 3 B_1(r_s),
\tag{19}
$$

a value which assumes complete isotropization of the relativistic particles' momentum distributions in the downstream region. Figs. 5-8 show $B_2$ for PSR B1259–63 defined in Eq. 18 and 19 as a function of orbital phase $\theta = \phi + \pi$ for different values of the outflow parameter $\Upsilon$.

---

[3]This might have been the case for a Crab-like pulsar embedded in a Be star wind with $\dot{M} \sim 10^{-8}$ $M_\odot$. No such system is known today, and the spindown luminosity of PSR B1259– 63 is more than two orders of magnitude less than that of the Crab pulsar.

### 5.1. Pulsar Cavity Results

We now give a few relevant results for the pulsar cavity characteristics as a function of the outflow parameter $\Upsilon$. For simplicity we consider here coplanar models, and we fix the outflow index $n$ and initial outflow velocity $v_0$ to their best fit values (see below), $n = 2.5$, $v_0 = 10^7 \, \mathrm{cm \, s^{-1}}$. We consider values of $\Upsilon = 10^1 - 10^3$ spanning the range from reasonably low to high momentum fluxes in the Be star wind (e.g., W88). Larger values of $\Upsilon$ would require peculiar outflow conditions for SS 2883 and they would be in contradiction with the observed low optical depth of cold nebular material determined by X-ray observations. We also fix, following the analysis of Paper II, the values of the pulsar wind parameters to be $\gamma_1 = 10^6$ and $\sigma = 0.02$.

*Case with $\Upsilon = 10$*

This is the case of a relatively low mass outflow rate from the Be star surface and a low outflow pressure. In this case, the shock radius is expected to be quite close to the Be star surface and the circumstellar disk can be disrupted by the pulsar's dynamic pressure. Fig. 5 gives the values of the shock radius, ions' Larmor radius $R_L$, compressed magnetic field $B_2(r_s)$, and other radiation timescales (discussed below) for the case characterized by an outflow parameter $\Upsilon = 10$. As expected, this case is characterized by a large IC cooling near periastron of the pulsar/outflow region, and Fig. 5 shows that even the IC cooling timescale in the relativistic regime becomes smaller than all other relevant timescales near periastron. The relativistic inverse Compton timescale reaches $\tau_{icr} \sim 200$ s, i.e., a factor of $\sim 3$ times smaller than the timescale for adiabatic expansion in the pulsar cavity $\tau_{ad}$. Strong IC cooling implies a substantial high-energy flux in the GeV-TeV energy range that would be difficult to reconcile with the spectral cutoff near $\sim 1$ MeV observed near periastron (T96, see also Paper II). For this reason, outflow models of SS 2883 with a low value of $\Upsilon$ are not supported by high-energy observations.

*Case with $\Upsilon = 100$*

This is the case for intermediate values of the mass outflow parameter, $\Upsilon = 10^2$. The shock radius is in this case further out from the surface of the Be star compared to the case with $\Upsilon = 10$. The compressed magnetic field strength is correspondingly larger than the previous case being of order of $\sim 1$ G near periastron. The circumbinary disk can be only partially disrupted by the pressure of the MHD pulsar wind. The shock radius $r_s$ can reach a value of $\sim 3 \cdot 10^{12}$ cm at periastron. Fig. 6 shows the relevant quantities, and we obtain that $\tau_{icr} \sim 500$ s and larger by a factor of $\sim 2$ than the adiabatic timescale $\tau_{ad}$. Synchrotron and relativistic IC cooling timescales turn out to be of similar values throughout the PSR B1259–63 orbit.

*Case with $\Upsilon = 10^3$*

This is the case with relatively large mass outflow parameter, i.e., mass loss rate combined with large outflow velocity. The shock radius is now typically $\sim 1/10$ of the orbital distance ($r_s \sim 10^{12}$ cm at periastron), and the circumbinary Be star disk is only marginally perturbed by the pulsar cavity. IC cooling in the relativistic regime is in this case negligible throughout the PSR B1259−63 orbit, and cooling efficiently takes place by synchrotron and IC cooling in a moderately relativistic regime (see Fig. 8).

## 6.   Accretion vs. non-accretion processes in the PSR B1259−63 system

We are now ready to confront the first important issue regarding the origin of high-energy emission from the PSR B1259−63 system, i.e., whether gravitationally-driven processes (accretion onto the neutron star surface or mass capture near pulsar magnetosphere in a propeller regime) play a relevant role. Previous investigations (carried out without knowledge of the recent results of the multiwavelength campaign near periastron) have discussed this issue in general for a variety of assumptions. Kochanek (1993) discussed the pulsar/outflow interaction in terms of a hydrodynamical shock allowing in principle a pulsar wind/ouflow coupling $\xi_p$ parameter to vary in a broad range $10^{-5} − 1$. A bremsstrahlung contribution to the X-ray luminosity from the heated mass outflow was estimated to be $\sim 10^{31}$ erg s$^{-1}$, with no detailed calculation of the intensity and spectrum expected from the pulsar wind contribution. King and Cominsky (1994, herafter KC94) discuss the pulsar/outflow interaction near *apastron* ignoring the dynamical effect of pressure originating from the pulsar wind (effectively assuming $\xi_p \ll 10^{-4}$ and a quasi-stable equilibrium at a boundary region outside the pulsar light cylinder radius). Gravitationally-driven accumulation of material near the magnetospheric boundary of PSR B1259−63 of radius $r_M$ is claimed by KC94 to be responsible for the X-ray emission near apastron as observed by ROSAT (CRJ94; see, however, Greiner *et al.* 1995). The resulting X-ray luminosity would be

$$L_{X,M} \simeq \frac{G \, M_{pulsar}}{r_M} \dot{M}_a \qquad (20)$$

where $G$ is Newton's constant, $M_{pulsar}$ the pulsar mass, and $\dot{M}_a$ the mass accumulation rate[4] given by $\dot{M}_a \simeq (R_{grav}/d)^2 \, \dot{M}$ where the Bondi gravitational capture radius is defined as

$$R_{grav} = \frac{2 \, G \, M_{pulsar}}{(\mathbf{v}_w - \mathbf{v}_p)^2 + c_s^2} \qquad (21)$$

with $c_s$ the gas sound speed. Since radio pulsations are not quenched at apastron, KC94 assume $r_M \sim R_{lc} \simeq 2.2 \cdot 10^8$ cm.

In an entirely different approach, TAK94 carried out a calculation of the pulsar cavity evolution of PSR B1259−63 for the physically meaningful assumption of $\xi_p \sim 1$. Their main conclusion is that accretion or magnetospheric propeller effect near periastron would imply a value of the mass outflow parameter larger than $\Upsilon \sim 10^3$. Further investigations of the propeller effect in the PSR B1259–63 system with the assumption $\xi_p \sim 1$ were carried out by Campana *et al.* (1995) and Ghosh (1995), who confirmed the TAK94 analysis concluding that the pulsar wind pressure barrier cannot be overcome in the PSR B1259–63 system near periastron unless the Be star mass loss rate is very large. Ghosh (1995) assumes in his calculations of the pulsar torque a large value of the Be disk outflow parameter $\Upsilon \gtrsim 10^4$. M95 reiterated a low-$\xi_p$ model of interaction in interpreting pulsar timing data near periastron passages of PSR B1259-63.

---

[4]We note here that the KC94 terminology emphasizing 'magnetospheric accretion' is ambiguous and should be avoided.

We have available today detailed information on the X-ray and soft $\gamma$-ray emission from the PSR B1259–63 system, and most of the previously proposed models of pulsar/outflow interaction can be shown to be in disagreement with observations. The relatively low X-ray luminosity of the system near periastron, its unpulsed nature, the constant and very low value of the absorption column density throughout the PSR B1259–63 orbit all argue against accretion of material onto the neutron star surface near periastron. For example, the characteristics of PSR B1259–63 near periastron are in sharp contrast with those of the transient accreting system A0538-66 during its high state (Skinner *et al.* 1981). Table 2 summarizes the observed features for the PSR B1259–63 system and A0538-66 in its high and low states. Since the geometry of pulsar/outflow interaction near periastron may be similar for PSR B1259–63 and A0538-66, Table 2 suggests to us that overcoming the pulsar wind's dynamic pressure is most likely to require a large mass outflow rate and therefore a large X-ray luminosity.

Furthermore, and this is an important issue also for the discussion of a possible propeller effect, the fourth ASCA observation of the PSR B1259–63 system of Feb. 28 1994 (H96) shows features of the X-ray emission (intensity, spectrum, column density) which are similar to previous observations immediately before, during and after periastron. Since pulsed radio emission from PSR B1259–63 was clearly visible again in late February 1994 (M95, J96), whatever mechanism producing the X-ray emission in February 1994 is likely to be similar to the mechanism operating near periastron. Accretion onto the surface of the neutron star is clearly excluded from a phenomenological point of view. The discussion given immediately below focusses on the likelihood of the propeller effect and can be easily applied to the analysis of surface accretion.

The likelihood of the propeller effect in the PSR B1259–63 system near periastron deserves more discussion. Since the pulsar light cylinder radius is substantially larger than its corotation radius ($R_{cor} = 2.1 \cdot 10^7$ cm), any flow reaching $R_{cor}$ will necessarily quench the pulsar mechanism and its MHD wind. This gravitationally trapped gas might interact with the pulsar magnetosphere which is expected to be rotating too fast to allow accretion onto the neutron star surface for the conditions of the PSR B1259–63 system near periastron. The propeller effect can in principle lead to X-ray dissipation of the gas kinetic and potential energy with a resulting maximum luminosity given by Eq. 20 where $R_{cor} \lesssim R_M \lesssim R_{lc}$. Electromagnetic torques may operate on the neutron star and can contribute to its effective spindown (e.g., Ghosh 1995). However, reaching such a close distance to the pulsar for the Be star gaseous material is difficult if not impossible for the PSR B1259–63 geometry. It is customary to parametrize the likelihood of gravitational trapping of mass outflows in binary system by making use of the gravitational Bondi radius $R_{grav}$ defined in Eq. 21. A necessary (but not necessarily sufficient[5]) condition for reaching a small distance from the pulsar is therefore that the shock radius $r_s$ becomes less than the gravitational radius along the pulsar orbit, i.e.,

$$r_s < R_{grav} \qquad (22)$$

Satisfying Eq. 22 depends crucially on the characteristics pulsar/outflow interaction and Be star outflow parameters ($\Upsilon, v_0, n, f_w$). Eq. 22 turns out to be impossible to satisfy for the pulsar cavity models presented in Figs. 5- 7. Only models with $v_0 \sim 10^6 \, \mathrm{cm \, s^{-1}}$ or $\Upsilon \gtrsim 10^4$ can lead to accretion. We carried out extensive calculations of coplanar and misaligned models of the PSR B1259–63 system and concluded that models with $v_0 \sim 10^6 \, \mathrm{cm \, s^{-1}}$ are not in agreement with the detailed temporal behavior provided by the X-ray data (see Sect. 10 and Table 2. We are left with models characterized by $\Upsilon \gtrsim 10^4$ which also are not in agreement with X-ray data (see Table 2). We assume here that the observed X-ray/soft $\gamma$-ray properties of the PSR B1259–63 system near periastron are smooth and not subject to unobserved changes of luminosity or column density. It is therefore possible to state that propeller effect in the PSR B1259–63 system is unlikely because of two main reasons:

(1) the X-ray luminosity expected from Eq. 20 for a mass flow sufficiently strong to reach a $R_M$ between $R_{cor}$ and $R_{lc}$ is larger than the observed luminosity by one-two orders of magnitude (the continuous monitoring by OSSE during periastron ensures that no sudden flare of emission was produced during the periastron passage);

(2) mantaining a constant and very low column density of cold material $N_H$ throughout the *whole* PSR B1259–63 orbit turns out to be impossible to satisfy for Be star outflows with $\Upsilon \gtrsim 10^4$.

The first argument assumes that accreting gas accumulating at the magnetopause has a cooling time short compared to the free fall time at $R_m$. One can readily show that at accretion rates large enough to satisfy (22), bremsstrahlung cooling alone suffices to yield $L_X = GM_p\dot{M}/R_M$ (see Burnard *et al.* 1983, for a general discussion of the bremsstrahlung emission from matter accumulating at a magnetopause). Therefore, the discrepancy between the X-ray observations and the accretion rates needed to put PSR B1259–63 into the hypothesized propeller regime cannot be avoided by assuming a small radiative efficiency for the accreting gas.

Furthermore, the fourth ASCA observation in late February 1994 demonstrates that the radio pulsar mechanism and unpulsed X-ray emission coexist, as well as showing that at this epoch, the radio emission's properties are not different from other, electromagnetically isolated millisecond pulsars. Models hypothesizing the propeller effect clearly have great difficulty in explaining the radio emission characteristic of electromagnetically isolated pulsars while at the same time hypothesizing the drastic alteration in the magnetospheric electrodynamics imposed by inserting a shell of dense, nonrelativistic plasma inside the light cylinder. Just such a demand is made of propeller models for the fourth ASCA observation. A pulsar/outflow model capable of mantaining a very low column density and moderate X-ray luminosity throughout the PSR B1259–63 orbit is clearly needed. The pulsar wind/outflow interaction with $\xi_p \approx 1$ distances from the pulsar large compared to the light cylinder radius $R_L = 2,280$ km easily satisfies this requirement, and is the model that we will consider in the remainder of the paper.

---

[5]See, e.g., Brookshaw & Tavani (1995).

## 7. Radiative Regimes of the PSR B1259–63 System

In this section we summarize the general properties of the radiative timescales of the PSR B1259–63 system (TAK94).We showed in TAK94 that the PSR B1259–63 system offers a variety of different radiative regimes depending on the effectiveness of cooling and acceleration processes at the pulsar/outflow shock interaction site. We assume the relativistic termination shock of the pulsar wind has the ability to accelerate a non-thermal power law distribution of post-shock $e^{\pm}$-pairs. Furthermore, in order to give a specific time scale for reference, we normalize the acceleration rate to the gyration frequency of relativistic ions encountering the shocked magnetic field of the pulsar wind, since particle-in-cell simulations carried out by HAGL92 showed that a relativistic shock wave energetically dominated by heavy ions indeed efficiently converts flow energy into power law spectra of positrons through cyclotron absorption of the large amplitude magnetosonic waves generated *within* the shock structure. Further unpublished quasi-linear calculations confirm that the acceleration rate is of this order of magnitude for both positrons and electrons. Therefore, the acceleration time scale is scaled to[6]

$$\tau_{\rm acc} \sim \Omega_i^{-1} = \frac{m_i c \gamma_1}{Z e B_2} =$$
$$\simeq 200 \left(\frac{\sigma}{0.02}\right)^{-1/2} \left(\frac{(1+\sigma)}{n_{10} v_7^2}\right)^{1/2} \left(\frac{\dot{P}_0}{P_{47}^3}\right)^{1/2} {\rm sec;} \qquad (23)$$

where $\Omega_i$ is the ion's relativistic cyclotron frequency at the shock front, $m_i$ and $Z$ the ion mass and charge, and $\dot{P}_0$ and $P_{47}$ the PSR B1259–63 spin time derivative and period, respectively. The shock acceleration mechanism produces a non-thermal downstream particle energy distribution which can be approximated as a power-law distribution of electron/positron energies $N(\gamma) \propto \gamma^{-p}$ for $\gamma_1 < \gamma < \gamma_m$ of index $p$. The maximum energy $\gamma_m$ depends on the model and radiative conditions.

---

[6]Our results do not exclude faster acceleration mechanisms with $\tau_{\rm acc}$ smaller than that of Eq. 23.

The overall efficiency of the shock in transforming the electromagnetic and kinetic energy of the pulsar wind is also another important feature of collisionless pulsar shocks. Observations and theory of the Crab nebula (see HAGL92 and GA94) give a fraction $\varepsilon_a \sim 0.2$ of the pulsar energy flux being transformed into the final energy of *nonthermal* $e^{\pm}$-pairs. So long as the acceleration time scale is short compared to the downstream flow $\tau_f$ and radiation cooling times $\tau_c(\gamma)$ for *all* energies $\gamma m_{\pm} c^2$ of the accelerated $e^{\pm}$, the same fraction of the pulsar's spindown luminosity may be assumed to go into the nonthermally accelerated pairs in the PSR B1259–63 system. The radiative conditions of the Crab nebula MHD shock differ from those of the PSR B1259–63 system in that the Crab nebula is a *diffuse plerion* (Arons & Tavani 1993, hereafter AT93), with $\tau_c \gg (\tau_{acc}, \tau_f)$ for almost all the accelerated particles ($\gamma_1 \leq \gamma \leq \gamma_m$). Therefore, the acceleration process at the shock front is complete before the radiation and adiabatic losses begin to extract energy from any of the $e^{\pm}$. If the radiation and magnetic field energy densities in the PSR B1259–63 system were as low as they are in the Crab Nebula, the maximum Lorentz factor of $e^{\pm}$-pairs accelerated at the shock front in the PSR B1259–63 system might reach $\gamma_m \simeq (m_i/m_{\pm})\gamma_1 \sim 2 \cdot 10^9$, comparable to the maximum energies inferred from the Crab Nebula's synchrotron emission. However, in the case of PSR B1259–63 , because of intense cooling of shocked relativistic particles by the combined effect of synchrotron radiation and inverse Compton scattering, $\tau_c(\gamma_m)$ might be comparable to or smaller than $\tau_{acc}$, in which case the $e^{\pm}$ cannot reach the highest energy $\gamma_m m_{\pm} c^2$ of which the acceleration mechanism (whatever it may be) is capable. The pulsar cavity of PSR B1259–63 resembles such a *compact plerion* (AT93) especially near periastron, where the compression of the pulsar cavity is expected to be larger and the radiative background field and local magnetic field energy is the largest. We will show below that from the shape of the power-law X-ray emission of the PSR B1259–63 system extending up to $\sim 200$ keV near periastron, we can obtain a lower limit on the ratio of the maximum downstream energy of accelerated pairs to $\gamma_1$, i.e., $\gamma_m/\gamma_1 \gtrsim 10$ (see the more extensive discussion of Sect. 10). We then obtain $\tau_c(\gamma_1) \gg \tau_{acc}$, and $\tau_c(\gamma)$ can become less than $\tau_{acc}$ only for particles whose energy is at least ten times $\gamma_1 m_{\pm} c^2$. In this circumstance, the efficiency $\varepsilon_a$ with which the pulsar creates nonthermally radiating particles is the same as in the diffuse plerions, although the efficiency with which the pulsar creates energy *stored* in relativistic $e^{\pm}$ may be reduced.

The dynamical flow time of particles streaming with velocity $V_f$ inside the pulsar cavity is

$$\tau_f \simeq 3\,\tau_{ad} \simeq 100 \quad \left(\frac{1}{n_{10} f_p v_7^2}\right)^{1/2} \frac{c/3}{V_f} \text{ sec,} \qquad (24)$$

where $\tau_{ad} = r_s/V_f$ is the adiabatic flow time to reach the 'apex' of the pulsar cavity (we have taken the post-shock flow velocity to be $V_f = c/3$). The factor of 3 in Eq. 24 takes into account the non-spherical shape of the pulsar cavity and the residence time of particle streaming out from the pulsar and hitting the shock region at different angles with respect to the 'apex' of the cavity. Since the flow and acceleration timescales near the PSR B1259–63 periastron can be comparable, $\tau_f \simeq \tau_{acc}$, adiabatic losses can in principle provide serious competition with the non-thermal $e^{\pm}$-pair acceleration process at the pulsar wind termination shock.

Radiation losses can also provide substantial competition with the acceleration of the pairs, especially to high energies $\gamma \gg \gamma_1$. Radiative losses due to synchrotron radiation of accelerated $e^{\pm}$-pairs in the local magnetic field at the shock and IC cooling of $e^{\pm}$-pairs in the background low-energy radiation field limit the maximum Lorentz factor obtainable at the shock. The synchrotron loss time $\tau_s$ can be approximated as

$$\tau_s(\gamma) = \frac{3m_{\pm}^3 c^5}{e^4 B_2^2 \gamma} \tag{25}$$

$$\simeq 3,600 \; \frac{1}{v_7^2 n_{10}} (1+\sigma) \left(\frac{\sigma}{0.02}\right)^{-1/2} \left(\frac{\dot{P}_0}{P_{47}^3}\right)^{-1/2} \frac{\gamma_1}{\gamma} \;\; \text{sec.} \tag{26}$$

The IC cooling timescale of high energy electrons against photons in the non-relativistic Thomson regime ($\gamma \, \epsilon \ll m \, c^2$) from the Be star is

$$\tau_{ic} = (\mathcal{E}_B/\mathcal{E}_R) \, \tau_s, \tag{27}$$

where $\mathcal{E}_B$ is the energy density of the post-shock electromagnetic field, and $\mathcal{E}_R$ is the energy density of the Be star radiation. The magnetic field energy density at $r_s$ is

$$\mathcal{E}_B(r_s) = \frac{B_2^2(r_s)}{8\pi}, \tag{28}$$

and the photon energy density (mostly IR-optical-UV) is given by[7]

$$\mathcal{E}_R(r_s) = \frac{R_*^2 \sigma_B T^4}{\overline{R}^2 c}, \tag{29}$$

where $\sigma_B$ is the Stefan-Boltzmann constant, $R_*$ is the radius of the Be star, $T$ is the Be star's effective temperature, and $\overline{R} \simeq d - r_s - R_*$ is the distance between the closest point on the surface of the Be star and the shock.

Eq. 27 is valid only in the Thomson regime. In case of a radiation bath at the photospheric temperature of SS 2883 ($T \simeq 27,000$ K $\simeq 2.3$ eV) this approximation is valid *only for* $\gamma_1 \lesssim 10^5$. For larger wind four velocities, Compton scattering occurs in the relativistic regime for which it is necessary to take into account the Klein-Nishina decrease of the cross section as a function of center-of-mass energy. The dimensionless parameter characterizing the scattering (Blumenthal and Gould, 1970; hereafter referred as BG) is

$$\Gamma_\epsilon = \frac{4 \, \gamma \, < \epsilon >}{m_{\pm} \, c^2} \tag{30}$$

---

[7]It can be shown *a posteriori* that the synchrotron self-Compton emission is negligible in the case of the PSR B1259–63 system. The dominant contribution to IC cooling originates from the surface or disk surroundings of the Be star companion.

with $<\epsilon>= 2.7\,k_B\,T$ the average photon energy of the radiation blackbody field ($k_B$ is Boltzmann's constant). In the case of PSR B1259–63, $\Gamma_\epsilon \sim 50\,\gamma_{1,6}$, where $\gamma_{1,6} = \gamma_1/10^6$. The IC scattering timescale in the relativistic regime $\tau_{icr}$ is (Paper II)

$$\tau_{icr} \simeq \frac{\gamma}{(70.8)\,\mathcal{E}_R(d-r_s)\,\mathcal{F}(\gamma)} \quad \text{s} \tag{31}$$

where $\mathcal{F}(\gamma) \simeq \pi^2/6[\ln(4\,\gamma\,k_B T/m_\pm\,c^2) - 1.41] - 0.3467$ (e.g., BG).

The total radiation rate per radiating $e^\pm$-pair of a given $\gamma$ at the shock $\tau_r^{-1}$ can be written as

$$\frac{1}{\tau_c'} = \frac{1}{\tau_s} + \frac{1}{\tau_{ic}'} \tag{32}$$

where $\tau_{ic}' = \min(\tau_{ic}, \tau_{icr})$ is the relevant IC timescale (either the non-relativistic value or the relativistic one). If $\tau_c$ is long compared to the flow time in the pulsar cavity $\tau_f$, the radiative losses are negligible, the flow is adiabatic and the resulting radiation nebula can be defined as *diffuse* (AT93). On the other hand, if the cooling timescale ($\tau_s$ or $\tau_c'$) is short compared to $\tau_f$, the shock acceleration of pairs produces a spectrum of $e^\pm$-pairs which is cut off at high photon energies, and a *compact* (AT93) nebula forms. If $\tau_c \ll \tau_{\rm acc}$, the incoherent radiation losses interfere with the acceleration process in the shock front itself, which produces a nebula which we call *ultra-compact*.

The maximum Lorentz factor $\gamma_m$ achievable in the complex radiative environment of the PSR B1259–63 system can therefore be obtained by the condition

$$\frac{\tau_c}{\tau_{\rm acc}} \sim \frac{1}{\gamma_m\,\gamma_1} \tag{33}$$

The value of $\gamma_m$ from Eq. 33 is substantially smaller than the value obtained for diffuse plerions such as the Crab Nebula. In the absence of strong IC up-scattering in the MeV-GeV energy range, we can therefore predict the existence of a spectral cutoff at intermediate $\sim$MeV photon energies for the resulting synchrotron emission (TAK94). Fig. 2 of TAK94 shows the behavior of the ratio $\tau_{\rm acc}/\tau_c$ as a function of orbital phase for different values of the outflow parameter $\Upsilon$ and for $n = 3$.

We can therefore consider the different radiation regimes by considering the set of ratios of timescales $\tau_{\rm acc}/\tau_c$, $\tau_f/\tau_c$, $\tau_{ic}'/\tau_s$ that are characteristics of the shock emission of the PSR B1259–63 system. TAK94 discuss the possible alternatives. The most notable feature of Fig. 2 of TAK94 is the possible relevance of non-relativistic inverse Compton scattering for the cooling process near periastron. For a broad range of pulsar and outflow parameters, the rate of non-relativistic inverse Compton cooling is several orders of magnitude larger than the synchrotron cooling rate near periastron if $\gamma_1 \lesssim 10^6$. Furthermore, the crucial condition to ensure non-thermal shock acceleration, i.e., $\tau_{\rm acc} \ll \tau_c$ may not be in principle satisfied. This 'inverse Compton catastophe' (see Paper II) can lead to the absence of non-thermal acceleration of particles at the shock, in contradiction with the observed power-law X-ray emission of the PSR B1259–63 system near periastron. On the other hand, near apastron and for a relatively large fraction of the orbit, the condition $\tau_{acc} \ll \tau_c$ can be more generally satisfied, and non-thermal particle acceleration can more easily occur. It is therefore possible that more than one radiative regime applies to the orbital motion of PSR B1259-63, for a fixed combination of Be star wind parameters.

### 7.1. Non-thermal diffuse nebular emission

This is the case characterized by $\tau_{acc} \ll (\tau_f, \tau_c)$, and $\tau_c \gg \tau_f$. This case applies to the part of the pulsar orbit far from the Be star (near apastron) for a broad range of input outflow parameters. Except for cases with $1 \lesssim \Upsilon \lesssim 10$, synchrotron cooling dominates the emission near apastron, i.e., near $\theta = 0$ (see Figs. 6-7). Since the flow time in the shock front is much less than the synchrotron loss time in the downstream flow, the flow is adiabatic both in the ions and in the pairs. The resulting synchrotron nebula at distances from the pulsar $r \gtrsim r_s$ is diffuse, $j_\varepsilon^{(n)} \propto \varepsilon^{-1/2}$ (energy/cm$^3$-sec-energy) with $\varepsilon$ the photon energy.

The energy range of the power-law spectrum in the diffuse nebular regime can be written as $\varepsilon_1^{(n)} < \varepsilon < \varepsilon_m^{(n)}$, where

$$\varepsilon_1^{(n)} = 0.3\gamma_1^2 \hbar\omega_c(B_2) \approx 1.8\,v_7 \left[\frac{n_{10}}{(1+\sigma)}\right]^{1/2} \left(\frac{\sigma}{0.02}\right)^{1/2} \text{ keV} \tag{34}$$

with $\omega_c$ the electron/positron cyclotron frequency, corresponds to the critical energy of the single particle synchrotron emissivity for pairs of energy $\gamma_1 m_\pm c^2$, and

$$\varepsilon_m^{(n)} = \gamma_{m,Crab}^2 \hbar\omega_c(B_2) \simeq 7\,v_7 \left[\frac{n_{10}}{(1+\sigma)}\right]^{1/2} \left(\frac{\sigma}{0.02}\right)^{1/2} \text{ GeV} \tag{35}$$

is the emission energy for the upper exponential cutoff of the synchrotron spectrum. Note that in Eq. 35 we have assumed for the maximum energy the value $\gamma_m$. This approximation is only valid if radiative losses can be neglected for the calculation of the post-shock energy distribution function over the *whole* particle energy range.

### 7.2. Non-thermal compact nebular emission

This is the case characterized by $\tau_{acc} \ll \tau_f, \tau_s, \tau_{ic}'$, and $\tau_s, \tau_{ic}' \leq \tau_f$ which applies to the pulsar passage near periastron. We distinguish two cases according to the nature of the cooling process of the PSR B1259–63 system.

### 7.2.1. Synchrotron losses dominate

Synchrotron losses may dominate if the only IC cooling occurs in the relativistic regime (for $\gamma_1 \gtrsim 10^6$) or, equivalently, if the shock radius $r_s$ is relatively close to the pulsar for $\Upsilon \gtrsim 10^2$. In this case, the range of the radiated photon spectrum is within the energies

$$\varepsilon_1 = 0.3\gamma_1^2\hbar\omega_c(B_2) \approx 3\,B_2\,(\gamma_1/10^6)^2 \text{ keV} \tag{36}$$

with $B_2$ in Gauss, and

$$\varepsilon_2 = \gamma_m^2\hbar\omega_c(B_2) \approx 1\,B_2\,(\gamma_1/10^6)^2(\gamma_m/10\gamma_1)^2 \text{ MeV}. \tag{37}$$

The synchrotron emission from a $e^\pm$-pair downstream energy distribution has therefore a 'knee' in the spectrum at $\epsilon_1$ and an exponential cutoff at $\epsilon_2$. Note that since

$$\epsilon_2/\epsilon_1 \sim (\gamma_m/\gamma_1)^2 \tag{38}$$

the observed X-ray and hard X-ray emission from the PSR B1259–63 system near periastron $\epsilon_2/\epsilon_1 \gtrsim 100$ corresponds to the constraint

$$(\gamma_m/\gamma_1) \gtrsim 10. \tag{39}$$

### 7.2.2. Inverse Compton losses dominate

If the pulsar wind of PSR B1259–63 is dominated by $e^\pm$-pairs with $\gamma_1 \sim 10^4 - 10^5$, IC scattering losses are likely to dominate the cooling near the periastron region for a large range of Be star outflow and pulsar wind parameters. In any case, IC scattering dominates the cooling if the shock radius $r_s$ is close to the surface of the Be star. In this case, the typical emitted photon spectrum will be approximately 'flat' (for isotropic photon background in the rest frame of the scattering high energy electron/positron) with a spectral break at the energy

$$\epsilon_{ic} \sim\,<\epsilon>\,\gamma_1^2 \sim (10 \text{ GeV})\,\gamma_{1,5}^2 \tag{40}$$

for non-relativistic scattering and

$$\epsilon_{icr} \sim\,<\epsilon>\,m\,c^2\,\gamma_1 \sim (1 \text{ TeV})\,\gamma_{1,6} \tag{41}$$

for scattering in the Klein-Nishina regime. The lack of observed emission in the MeV-GeV energy range and the relatively steep power law nature of the X-ray emission argue against substantial IC scattering cooling (see Paper II).

### 7.3. 'Thermal' nebular emission

This is the case characterized by $\tau_{\mathrm{acc}} \gg \tau_s, \tau_{ic}'$. If the shock-acceleration mechanism is not capable to produce a non-thermal post-shock $e^{\pm}$-pair energy distribution, the resulting high energy emission will reflect the relativistic Maxwellian shape of the injection. This mode of emission could be applicable to pulsar binaries characterized by very strong radiation cooling near periastron. A crucial test to shock-acceleration theory (see Paper II) can be provided by such a system. However, since the observed X-ray emission from the PSR B1259–63 system is consistent with power law emission from 1 keV to $\sim 200$ keV, we exclude this mode of emission in the case of the PSR B1259–63 system.

## 8. Shock Emission Luminosity

The intensity of the high energy emission from the interaction of PSR B1259–63 with the mass outflow of its Be star companion strongly depends on the local characteristics of the magnetic and IR, optical and UV energy density at the shock. The general behavior of the emission is that the intensity from synchrotron emission is expected to be relatively larger for shock radii $r_s$ near the pulsar (and with a correspondingly larger local magnetic field strength). Alternately, if the shock region is relatively close to the surface of the Be star, inverse Compton scattering dominates the cooling of the relativistic pairs at the shock, and therefore the high energy emission. Given the variety of possibilities for the characteristics of the Be star mass outflow, we have to consider different sequences of radiating 'regimes'. In general the high energy emission is expected to be time variable as a function of orbital phase, with both total intensity and spectrum varying.

The high energy shock luminosity $L_s$ depends ultimately on the ratio $\tau_{\mathrm{f}}/\tau_c$. If the cooling processes occur on a time scale that is short compared to the dynamical flow time scale $\tau_{\mathrm{f}}$, then

$$L_s \simeq \varepsilon_a \dot{E}_R F(\Omega_*, f_p), \qquad (42)$$

where $F(\Omega_*, f_p)$ is the fraction of the pulsar's wind which is stopped by the obstacle formed by the Be star's outflow, and $\varepsilon_a$ is the efficiency of pulsar wind energy conversion into accelerated particles. In the case of the Crab nebula, $F(\Omega_*, f_p) \simeq 1$, since the pulsar wind is completely halted by the supernova remnant. In the case of the PSR B1259–63 pulsar/outflow interaction, the quantity $F(\Omega_*, f_p)$ can differ substantially from unity. High energy observations of the Crab nebula (Clear *et al.* 1987; Nolan *et al.* 1993) and theoretical calculations (HAGL92) give $\varepsilon_a \sim 0.1 - 0.2$. As shown in Figs. 5-6 the case with $\tau_{\mathrm{f}}/\tau_c \gg 1$ can be realized near periastron for a large variety of input outflow parameters.

For $\tau_c \gtrsim \tau_{\mathrm{f}}$, we must consider the details of the radiation processes in order to determine the bolometric emission. Since most of the shock emission comes from the highest energy particles, the nonthermal part of the pair spectrum dominates the emission. Therefore, we assume for the $e^{\pm}$-pair number density function, $N_{\pm}(\gamma) \propto \gamma^{-p}$, with $p$ the exponent of the nominal post-shock energy distribution function undisturbed by cooling processes (as we will show in the next Section, cooling processes do affect the high-energy power law). For a pulsar cavity radiating volume $V_p \equiv (k\, r_s)^3$ with $k \sim 3$ since the shocked, adiabatically flowing particles fill a volume comparable to that of the whole cavity, and a cavity residence time $k\, r_s/V_f = \tau_f$, we can easily obtain the total energy of $e^\pm$-pairs in the radiating volume (see also AT93)

$$E_\pm = \epsilon_a \dot{E}_R \frac{k\, r_s}{V_f} = V_p \int_{\gamma_1}^{\gamma_m} N(\gamma)\, \gamma\, m\, c^2\, \mathrm{d}\gamma = V_p\, n_\pm m\, c^2\, \mathcal{Q}(s, \gamma_1, \gamma_m) \qquad (43)$$

where we the fraction $\epsilon_a$ of the pulsar energy loss goes into the non-thermal component of the $e^\pm$-pair energy distribution function. Here

$$\begin{aligned}
\mathcal{Q}(s, \gamma_1, \gamma_m) &= \int_{\gamma_1}^{\gamma_m} \gamma^{1-p}\, \mathrm{d}\gamma \\
&= (p-1)\, \gamma_1^{p-1} \frac{1}{1 - (\gamma_1/\gamma_m)^{p-1}} \ln\left(\frac{\gamma_1}{\gamma_m}\right) \quad \text{if p = 2} \\
&= \frac{p-1}{p-2}\, \gamma_1^{p-1} \frac{1}{1 - (\gamma_1/\gamma_m)^{p-1}} \left(\frac{1}{\gamma_1^{p-2}} - \frac{1}{\gamma_m^{p-2}}\right) \quad \text{if p} \neq 2 \qquad (44)
\end{aligned}$$

We can therefore express the total number of shocked $e^\pm$-pairs for $p \neq 2$ as

$$(k\, r_s)^3 n_\pm^{nt} = \frac{k r_s}{V_f f_p} \frac{\epsilon_a\, \dot{E}_R}{\gamma_1 m_\pm c^2} \frac{(p-2)(1 - x^{p-1})}{(p-1)(1 - x^{p-2})}. \qquad (45)$$

where we defined $x \equiv \gamma_1/\gamma_m$. For $p = 2$ (case relevant for the PSR B1259–63 system) we obtain

$$(k\, r_s)^3 n_\pm^{nt} = \frac{k r_s}{V_f f_p} \frac{\epsilon_a\, \dot{E}_R}{\gamma_1 m_\pm c^2} \frac{(1 - x)}{\ln(1/x)}. \qquad (46)$$

Evaluating Eq. 46 for $x = 10^{-1}$ and $V_f = c/3$ we obtain

$$n_\pm^{nt} \simeq (0.4\, \mathrm{cm}^{-3})\, \epsilon_a\, \gamma_{1,6}^{-1}\, k^{-2}\, r_{s,13}^{-2} \qquad (47)$$

where $\gamma_{1,6} = \gamma_1/10^6$ and $r_{s,13} = r_s/(10^{13}\, \mathrm{cm})$. The $e^\pm$-pair injection rate from PSR B1259–63 is

$$\dot{N}_\pm = \frac{4\pi}{k^2} \frac{\epsilon_a}{V_f/c} \frac{1 - x}{\ln(1/x)} \dot{N}'_\pm \approx 0.3 \left(\frac{3}{k}\right)^2 \frac{\epsilon_a}{0.2} \frac{c/3}{V_f} \dot{N}'_\pm, \qquad (48)$$

where $\dot{N}'_\pm \equiv \dot{E}_R/(\gamma_1 m_\pm c^2) \simeq 9.8 \cdot 10^{35}\, \mathrm{s}^{-1}$ is the upper limit to the pair injection rate which would be the case if all the wind's energy were in the kinetic energy of the pairs. The numerical value assumes an acceleration efficiency comparable to that of the shock in the Crab Nebula. In general the shock emissivity is a function of the shape of the post-shock particle energy distribution and it is expected to change in time following the 'compression' (i.e., a change of $r_s$) of the pulsar cavity, the time change of the particle distribution function and the radiative regime. The time evolution of the particle distribution function is discussed in Sect. 9.

For shock-driven synchrotron emission, the total bolometric luminosity of the flowing plasma can be written as

$$L_{s,sync}(r_s) = \epsilon_s \, V_p(r_s) \, \left[ n_+^{nt}(r_s) + n_-^{nt}(r_s) \right] \, \frac{4}{3} \sigma_T \, c \, \frac{B_2(r_s)^2}{8\,\pi} \int_{\gamma_1}^{\gamma_m} f'(\gamma) \, \gamma^2 \, \mathrm{d}\,\gamma \qquad (49)$$

where $\epsilon_s$ is the efficiency of synchrotron cooling withing the flow time, $\epsilon_s = (1 + \tau_s/\tau_f)^{-1}$, and $f'(\gamma)$ the *normalized* non-thermal particle distribution of post-shock $e^{\pm}$-pairs defined by $N_{\pm}(\gamma) = n_{\pm}^{nt} \, f'(\gamma)$, with $f'(\gamma) = (p-1) \, \gamma_1^{p-1} \, [1 - (\gamma_1/\gamma_m)^{p-1}]^{-1} \, \gamma^{-p}$. Note that $L_{s,sync}(r_s)$ in first approximation depends on $r_s$ as

$$L_{s,sync}(r_s) \sim \tau_f \, B_2^2 \sim \frac{1}{r_s} \qquad (50)$$

We can therefore obtain an useful expression for $L_{s,sync}(r_s)$

$$L_{s,sync}(r_s) \simeq \frac{3}{16\,\pi} \, \frac{\epsilon_s \, \epsilon_a \, \dot{E}_R}{m_{\pm} c^2} \, \frac{k \, r_s}{f_p V_f} \, c \, \sigma_T B_2^2 \gamma_m \mathcal{A}(p,x) \qquad (51)$$

with

$$\begin{aligned} \mathcal{A}(p,x) &= \frac{p-2}{3-p} \, x^{p-2} \, \frac{1 - x^{3-p}}{1 - x^{p-2}} \quad \text{for p} \neq 2 \\ &= \frac{(1-x)}{\ln(1/x)} \qquad \text{for p} = 2 \ . \end{aligned} \qquad (52)$$

The non-relativistic IC luminosity $L_{s,ic}(r_s)$ of the shock is easily obtained from Eq. 49 with the change of magnetic field local energy density into the IR-optical-UV background photon energy density, i.e.,

$$L_{s,ic}(r_s) = \epsilon_{ic} V_p(r_s) \, \left[ n_+^{nt}(r_s) + n_-^{nt}(r_s) \right] \, \frac{4}{3} \sigma_T \, c \, \mathcal{E}_R \int_{\gamma_1}^{\gamma_m} f'(\gamma) \, \gamma^2 \, \mathrm{d}\,\gamma \qquad (53)$$

where $\epsilon_{ic} = (1 + \tau_{ic}/\tau_f)^{-1}$ is IC radiation efficiency of the pulsar cavity in the Thomson regime.

The energy loss in the relativistic IC regime occurs with discontinuous energy changes of the scattering $e^{\pm}$-pairs in the background photon field. The discontinuous energy loss rate in the relativistic regime $\gamma_{icr}$ can be obtained (e.g., BG) and the luminosity can be expressed as

$$L_{s,icr}(r_s) = \epsilon_{icr} V_p(r_s) \, \left[ n_+^{nt}(r_s) + n_-^{nt}(r_s) \right] \, m_{\pm} \, c^2 \int_{\gamma_1}^{\gamma_m} f'(\gamma) \, |\dot{\gamma}_{icr}| \, \mathrm{d}\,\gamma \ , \qquad (54)$$

with the average $\dot{\gamma}_{icr}$ given by

$$- \dot{\gamma}_{icr} \simeq (70.8 \, \mathrm{s}^{-1}) \, \mathcal{E}_R (d - r_s) \, \frac{1}{(\Theta/4.5 \cdot 10^{-6})^2} \int_0^{\infty} \frac{y}{e^y - 1} \left[ \ln(4 \, \gamma \, y \, \Theta) \, - \frac{11}{6} \right] \mathrm{d}\,y \ . \qquad (55)$$

The radiation efficiency of the pulsar cavity $\epsilon_{icr} = (1 + \tau_{icr}/\tau_f)^{-1}$ refers to the high-energy part of the IC spectrum in the relativistic Klein-Nishina regime. As we will see in the next sections, $L_{s,icr}(r_s)$ is relatively small for the PSR B1259–63 system compared to $L_{s,sync}$ as expected for a post-shock particle energy distribution initially peaked at $\gamma_1 \gtrsim 10^6$. Even adiabatic and cooling processes within the pulsar cavity timescale of the PSR B1259–63 system do not change the peak energy of the particle distribution function by more than a factor of $\sim 10$ (see next Section). We will see that the relevant IC cooling in the PSR B1259–63 system near periastron occurs in a moderately relativistic regime. The form of the spectrum depends on whether the radiation loss timescale for pairs with energy $\gamma = \gamma_1$ is short or long compared to the flow time. In general, higher energy particles can have radiation time short compared to the flow time, while the lower energy particle can have long radiative lifetime and therefore radiate inefficiently. This possibility forces us to consider the evolution of the particle spectrum in some detail.

## 9. Evolution of the particle energy distribution and radiation spectrum

Since the pulsar cavity (and therefore the non-thermal acceleration region) of PSR B1259–63 is subject to strong radiative cooling within the flow timescale of $e^\pm$-pairs, it is necessary to calculate the temporal evolution of the energy distribution function. The general solution an arbitrary pulsar wind particle energy distribution is a complex function of initial parameters. We will assume (from the results of Paper II) that IC scattering occurs in the moderately relativistic regime and that ultrarelativistic IC cooling of particles does not play a substantial role for the PSR B1259–63 system. It is also important to note that in all relevant models we will consider the flow time exceeds the acceleration time of the pairs for both shock-drift and resonant magnetosonic acceleration mechanisms. Therefore, we can consider the particles as injected at the shock with a fully formed power law distribution at energies $\gamma > \gamma_1$. They are subject to relaxation due to synchrotron, IC and adiabatic losses as they flow downstream. We treat the relativistic shock of pulsar wind particles as collisionless. Following a streamline, the age and loss equation for the pairs reads

$$\frac{\partial n}{\partial t} + v\frac{\partial n}{\partial s} + \frac{\partial}{\partial \gamma}(\dot{\gamma} n) = q(\gamma, s, t) - \frac{n}{t_f}, \qquad (56)$$

where $t_f = L/v$ is the flow time for the pair plasma of velocity $v$ to leave the "cometary" nebula whose length is $L$, $s$ is the distance along the streamline, and

$$\dot{\gamma} = -\frac{\gamma}{t_{ad}} - \frac{4\,\sigma_T}{3\,m_\pm\,c}\left(\mathcal{E}_B + \frac{\mathcal{E}_R}{\alpha_{ic}}\right)\gamma^2 = -\frac{\gamma}{t_{ad}} - \frac{\gamma^2}{T_c}\,. \qquad (57)$$

(we have omitted the explicit dependence of all quantities of Eq. 57 to the streamline distance $s$), where $t_{ad}$ is the adiabatic loss time,

$$t_{ad} \equiv \frac{r_s}{v\,k'} \qquad (58)$$

with $k' = 1, 2, 3$ for 1D, 2D and 3D adiabatic losses respectively, while

$$T_c^{-1} \equiv T_s^{-1} + T_{ic}^{-1} \qquad (59)$$

where

$$T_s \equiv \frac{6\pi m_\pm c^2}{c \sigma_T B^2}, \qquad T_{ic} \equiv \frac{3 m_\pm c}{4 \sigma_T \mathcal{E}_R} \alpha_{ic} \qquad (60)$$

are the synchrotron and IC loss time constants, respectively. These constants are related to the synchrotron and IC timescale for particles of energy $\gamma m_\pm c^2$ by $\tau_s = T_s/\gamma$, and $\tau_{ic} = T_{ic}/\gamma$. The constant $\alpha_{ic}$ of Eq. 57 (which turns out to be typically of order 5-10 for PSR B1259–63 parameters) takes into account deviations of the IC scattering process from the non-relativistic regime ($\alpha_{ic} = 1$ for Thomson scattering). Most of the observable radiation from the PSR B1259–63 pulsar cavity is produced by $e^\pm$-pairs advected away from the shock region whose energy has been substantially decreased by cooling and adiabatic losses. Since $\gamma \tau_{ad}/T_c \lesssim 1$ for the periastron parameters (see Sect. 10), anisotropic pitch angle scattering does not play a major role for the relevant $\gamma$'s. We therefore assume isotropic pitch angle scattering for the parameters of the PSR B1259–63 system.

We consider the simplest relevant solution of Eq. 57. Suppose $B$ and $v$ are constant along a streamline. Then solving the characteristic equation $d\gamma/dt = \dot{\gamma}$ yields the particle motion in energy space, with the result

$$\gamma(\gamma_0, t) = \gamma_0 \frac{e^{-\mu t/t_f}}{1 + \frac{t_{ad}\gamma_0}{T_c}(1 - e^{-\mu t/t_f})}, \qquad (61)$$

where we defined

$$\mu \equiv t_f/t_{ad}$$

with $\gamma_0$ a particle's energy at the shock, and $t$ the time following a fluid element along a streamline. It is important to realize that both synchrotron and IC cooling may affect the same radiating particles even though the resulting emission of synchrotron and IC cooling can be radiated in quite different energy bands. In the case of the PSR B1259–63 system, we find that most of the synchrotron emission is produced by moderately cooled $e^\pm$-pairs of energy $\gamma \sim 10^5$ and relative 'residence timescale' in the pulsar cavity near $\mu \sim 3$. The same population of cooled $e^\pm$-pairs will therefore lead, for the relevant parameters of the PSR B1259–63 system, to (observable) synchrotron emission in the keV-MeV band and (unobservable) IC emission above 10-100 GeV. Note that also low-energy particles in the 'thermal' part of the post-shock energy distribution follow the time evolution of Eq. 61. It turns out that these particles contribute to synchrotron emission below $\sim 1$ keV and to an IC emission (near 10 GeV) too weak to be detected with current instruments. In the following we focus on the energetic non-thermal component of the particle energy distribution.

Eq. 57 can be rewritten in a Lagrangian form, and following a particle orbit in time position and energy for a constant velocity, $s(t) = vt$, we obtain

$$\frac{d}{dt} n[\gamma(\gamma_0, s(t), t] + \frac{n[\gamma(\gamma_0, t), s(t), t]}{t_f} - \frac{n[\gamma(\gamma_0, t), s(t), t]}{t_{ad}} \left[ 1 + \frac{2 t_{ad}}{T_c} \gamma(\gamma_0, t) \right] = q[\gamma(\gamma_0, t), s(t), t].$$
(62)

We can integrate Eq. 62 by making use of Eq. 61 and obtain

$$
\begin{aligned}
n[\gamma(\gamma_0, t), t] &= e^{(\mu-1)t/t_f} \left[ 1 + \frac{t_{ad}\gamma_0}{T_c} \left( 1 - e^{-\mu t/t_f} \right) \right]^2 \\
&\times \int_0^t dt' \frac{e^{-(\mu-1)t'/t_f} q[\gamma(t'), t']}{\left[ 1 + \frac{t_{ad}\gamma_0}{T_c} \left( 1 - e^{-\mu t'/t_f} \right) \right]^2}
\end{aligned}
$$
(63)

For an efficient acceleration with acceleration timescale $t_a$ smaller than all other timescales, i.e., $t_a \ll t_f$, $t_{ad}$ and $t_s$ for all relevant energies, the injection layer can be approximated as 'thin' in energy space and $q = Q(\gamma_0)\delta(t)$. Then

$$n(\gamma(\gamma_0, t), t) = e^{(\mu-1)t/t_f} \left[ 1 + \frac{t_{ad}\,\gamma_0}{T_c} \left( 1 - e^{-\mu t/t_f} \right) \right]^2 Q(\gamma_0).$$
(64)

Now suppose that the acceleration mechanism produces a post-shock particle energy distribution function for $\gamma_{0,min} \lesssim \gamma_0 \lesssim \gamma_{0,max}$ of the form

$$Q(\gamma_0) = K \gamma_0^{-p}$$
(65)

with $p$ the exponent of the distribution and $K$ a constant. We can invert Eq. 61 and obtain

$$\gamma_0 = \frac{\gamma}{1 - \frac{t_{ad}\gamma}{T_c}(1 - e^{-\mu t/t_f})}.$$
(66)

By using Eqs. 65-66, Eq. 64 becomes a Kardashev-type (Kardashev, 1962) relation for the radiatively relaxed distribution function

$$n(\gamma, t) = K \gamma^{-p} \exp\{-[\mu(p-1)+1]/t_f\} \left[ 1 - \frac{t_{ad}\gamma}{T_c} \left( e^{\mu t/t_f} - 1 \right) \right]^{(p-2)},$$
(67)

Note that for $p < 2$ Eq. 67 shows the characteristic pile up of the relaxed distribution function at the energy $\gamma_c = (T_c/t_{ad})/(e^{\mu t/t_f} - 1)$. For $t \ll t_{ad}$ we obtain the familiar relation $\gamma_c = T_c/t$ (Kardashev, 1962). It is important to note that the relaxed distribution of Eq. 67 is cutoff for $\gamma > \gamma_c$ (except for the relevant case with $p = 2$). There is a relation (valid for $p \neq 2$ but that we will assume to hold also for the idealized degenerate case with $p = 2$) between the maximum advection time $T(\gamma)$ and the particle energy $\gamma$, of the type

$$\frac{T(\gamma)}{t_{ad}} = \ln \left( \frac{T_c}{t_{ad}\,\gamma} + 1 \right)$$
(68)

Note that the quantity $T(\gamma)$ is related with a position $L(\gamma)$ along the streamline where the radiation corresponding to a particular energy $\gamma$ is emitted, $L(\gamma) = T(\gamma)/v$ (typically $v = c/3$). We deduce that emission of (higher) lower energy is emitted relatively (close) far from the apex of the pulsar cavity.

We are now ready for the calculation of the high-energy spectrum of an unresolved source such as the PSR B1259–63 system. Let the photon emissivity (photons cm$^{-3}$ s$^{-1}$ keV$^{-1}$) be

$$j_\varepsilon(t) = \int d\gamma \, \frac{P_\varepsilon^s + P_\varepsilon^{ic}}{\varepsilon} \, n(\gamma, t),$$  (69)

where $P_\varepsilon^s$ is the synchrotron emissivity that we approximate as (e.g., Felten and Morrison, 1966)

$$P_\varepsilon^s = c\sigma_T \frac{B^2}{6\pi} \gamma^2 \delta(\varepsilon - \varepsilon_s),$$  (70)

with

$$\varepsilon_s = \varepsilon_c \gamma^2$$  (71)

and

$$\varepsilon_c = \frac{\hbar e B}{m_\pm c}.$$  (72)

and $P_\varepsilon^{ic}$ is the IC emissivity that can be approximated as (e.g., Felten and Morrison, 1966)

$$P_\varepsilon^{ic} = \frac{4}{3}\sigma_T c \, \frac{\mathcal{E}_R}{\alpha_{ic}} \gamma^2 \, \delta(\varepsilon - \varepsilon_{ic}),$$  (73)

with $\varepsilon_{ic} = <\varepsilon> \gamma^2$. Notice that only synchrotron emission contributes in the keV-MeV energy range for a cooled distribution of particles in the energy range $\gamma \sim 10^5 - 10^6$. The IC contribution to the emission is radiated in the $\sim 100$ GeV-TeV energy range for the choice of best fitting parameters of the PSR B1259–63 pulsar wind, and it is therefore consistent with the lack of emission in the 100 MeV-10 GeV as determined by EGRET (T96). However, it is important to realize that Eq. 69 is valid in general, and that for a different choice of pulsar wind parameters (for example a starting $\gamma_1 \sim 10^5$ or lower) would have caused a substantial contribution of IC emission in the EGRET energy range. Substituting Eq. 67 into Eq. 69 and integrating $j_\varepsilon$ over the whole volume of the particle flow, we finally obtain the photon spectral luminosity $L_\varepsilon$ (in photons s$^{-1}$ keV$^{-1}$) from the shocked wind for the special but most interesting case $p = 2$

$$L_\varepsilon = \frac{c\sigma_T B^2}{12\pi \varepsilon_c^2} \left(\frac{\varepsilon}{\varepsilon_c}\right)^{-(p+1)/2} \frac{AL}{\mu + 1} \left\{ 1 - \frac{1}{\left[1 + \frac{T_s}{t_{ad}}\left(\frac{\varepsilon_c}{\varepsilon}\right)^{1/2} + \frac{\tau'_{ic}}{t_{ad}}\right]^{(\mu+1)/\mu}} \right\}.$$  (74)

where $A$ is the projected area (along the flow) where most of the radiating particles of the relativistic shock are concentrated, $L$ the length of the downstream region, and $\tau'_{ic}$ the IC cooling timescale calculated for the same $e^{\pm}$-pair energy that gives rise to the synchrotron in the keV-MeV energy range. Note that in Eq. 74 we used Eq. 68. We can impose that the same $e^{\pm}$-pairs producing the synchrotron emission are those being cooled by IC in the moderately relativistic regime

$$\tau'_{ic} = T_{ic}\,(\varepsilon_c/\varepsilon)^{1/2}\,f(\phi) \tag{75}$$

where the function $f(\phi)$ represents absorption/reprocessing of the relevant soft energy photon background at the pulsar distance as a function of orbital phase. We model the absorption of optical/IR flux irradiating the pulsar cavity in the PSR B1259-63 system as

$$f(\phi) = \exp[+\phi^2/2\,\sigma_{ic}^2] \tag{76}$$

where $\sigma_{ic}$ is the angular size of the region more optically thin to the soft photon propagation from the Be star companion. Both thickness and reprocessing of optical/IR flux propagating through the Be star outflow and reaching the pulsar cavity critically affect $\tau'_{ic}$. Our parametrization of Eq. 76 takes into account: *(i)* the higher probability of IC cooling in the moderately relativistic regime near periastron, and *(ii)* the progressively more inefficient IC cooling as the pulsar more strongly interacts (splashing) with the Be star outflow at $\theta_1$ and $\theta_2$ and progressively recedes from the periastron region where IC scattering can effectively occur for $10^5 \lesssim \gamma \lesssim 10^6$. We have then two different radiation regimes:

**(A)** For photon energies such that $(T_s/t_{ad})(\varepsilon_c/\varepsilon)^{1/2} + \tau'_{ic}/t_{ad} \gg 1$, adiabatic losses dominate over synchrotron losses in the emitting particles, and the emitted spectrum is

$$L_\varepsilon \approx \frac{c\sigma_T B^2}{12\pi\varepsilon_c^2}\left(\frac{\varepsilon}{\varepsilon_c}\right)^{-(p+1)/2}\frac{AL}{\mu+1}\ . \tag{77}$$

**(B)** For a synchrotron/IC timescale substantially smaller than the adiabatic timescale, i.e., for $(T_s/t_{ad})(\varepsilon_c/\varepsilon)^{1/2} + \tau'_{ic}/t_{ad} \ll 1$, synchrotron/IC losses age the spectrum of the injected particles, sweeping the upper cutoff of the emitted photon spectrum through the observer's band. The results is the approximate spectrum

$$L_\varepsilon \approx \frac{c\sigma_T B^2}{12\pi\varepsilon_c^2}\left(\frac{\varepsilon}{\varepsilon_c}\right)^{-(p+2)/2}AvT_s \tag{78}$$

where, for simplicity, we assumed a particle relative residence time in the pulsar cavity $\mu$ substantially larger than unity.

Eqs. 77-78 constitute the most important result of our analysis of synchrotron/IC emission from the PSR B1259-63 system. The predicted photon spectral index of high-energy emission $\alpha = d\ln(L_\varepsilon)/d\ln(\varepsilon)$ (for the case $p=2$ relevant for the PSR B1259-63 system) is therefore a non-trivial function of $\mu$ and of the ratio $\frac{T_s}{t_{ad}}\left(\frac{\varepsilon_c}{\varepsilon}\right)^{1/2} + \frac{\tau'_{ic}}{t_{ad}}$

$$\alpha = -\frac{3}{2} - \frac{1}{2}\frac{\mu+1}{\mu}\frac{\frac{T_s}{t_{ad}}\left(\frac{\varepsilon_c}{\varepsilon}\right)^{1/2} + \frac{\tau'_{ic}}{t_{ad}}}{\left[1 + \frac{T_s}{t_{ad}}\left(\frac{\varepsilon_c}{\varepsilon}\right)^{1/2} + \frac{\tau'_{ic}}{t_{ad}}\right]^{1+(\mu+1)/\mu}}\frac{1}{1 - \left[1 + \frac{T_s}{t_{ad}}\left(\frac{\varepsilon_c}{\varepsilon}\right)^{1/2} + \frac{\tau'_{ic}}{t_{ad}}\right]^{-(\mu+1)/\mu}} \tag{79}$$

Eq. 79 can be applied for synchrotron emission in the keV-MeV energy range as long as the condition $\frac{T_s}{t_{ad}} \left( \frac{\varepsilon_s}{\varepsilon} \right)^{1/2} + \frac{\tau'_{ic}}{t_{ad}} < 1$ can be applied to $\varepsilon \gtrsim 1$ keV. From Eq. 49 we can obtain the integrated spectral luminosity (in erg s$^{-1}$) for a given energy band due to synchrotron emission is

$$L(x_1, x_2) = \frac{c \, \sigma_T \, v}{12 \, \pi} \frac{\mu}{1 + \mu} \, t_{ad}(r_s) \, B_2^2(r_s) \, I(\mu, x_1, x_2) \qquad (80)$$

where we have defined

$$I(\mu, x_1, x_2) = \int_{x_1}^{x_2} x^{-(p-1)/2} \left[ 1 - \frac{1}{[1 + (T_s/t_{ad}) \, x^{-1/2} + \frac{\tau'_{ic}}{t_{ad}}]^{(\mu+1)/\mu}} \right] \mathrm{d} \, x \qquad (81)$$

with $x = \varepsilon/\varepsilon_c$ and $x_1 = \varepsilon_1/\varepsilon_c$, $x_2 = \varepsilon_2/\varepsilon_c$, where $\varepsilon_1$ and $\varepsilon_2$ define the energy band of a particular detector. In the case of ASCA, we have $\varepsilon_1 = 1$ keV, and $\varepsilon_2 = 10$ keV, and for OSSE $\varepsilon_1 = 50$ keV, and $\varepsilon_2 = 200$ keV.

## 10.  Model calculations: comparison with observations

We can now compute the shock emissivity as a function of orbital phase by making use of the model presented in Sect. 9 and especially of Eqs. 79 and 80. It turns out that the time behavior of the X-ray luminosity and spectrum is the best diagnostic to discriminate between different pulsar/outflow interaction models. We consider variations of: *(i)* the Be star outflow overall geometry (coplanar to the pulsar orbit or misaligned), *(ii)* the outflow parameter $\Upsilon$ ($\Upsilon_1$ and $\Upsilon'_2$ for misaligned models), *(iii)* the index $n$, *(iv)* the IC cooling efficiency depending on the possible screening of IC cooling flux from the Be star surface as a function of outflow geometry and pulsar/outflow interaction.

We carried out an extensive series of calculations to probe the relevant parameter space for the PSR B1259–63 system. Table 3 gives a schematic summary of the main results, with a indication of the agreement or disagreement between calculated and observed X-ray and soft $\gamma$-ray emission.

*Misaligned models*

Figs. 8 and 9 show the results for the best model that represents the available X-ray data. The model is based on a *misaligned* geometry with a first 'splashing point' of the pulsar hitting the Be star equatorial wind at $\theta_1 = 120°$ (and a second point at $\theta_2 = \theta_1 + 180°$). These parameters are consistent with the pulsar orbital inclination of $\sim 35°$ deduced from radio observations (M95). The Be star disk opening angle at the pulsar orbital distance is assumed to be $\sigma_d = 30°$ (see Eq. 5 for definition of $\sigma_d$), the best outflow index is $n = 2.5$, and the magnitudes of the outflow in the 'polar' and equatorial regions are given by $\Upsilon_1 = 10^2$ and $\Upsilon'_2 = 7 \cdot 10^2$, respectively. It is important to emphasize that IC cooling of the particle energy distribution function as modelled in Sect. 9 is crucial to obtain both the 'double-humped' shape of the X-ray luminosity as well as the temporal behavior of the spectral index $\alpha$. We find that a variable 'screening' geometry is favored over models with uniform IC cooling photon irradiation of the pulsar cavity. Fig. 10 gives the calculated intensity and spectral index in (OSSE) the energy range 50-200 keV for the best misaligned model. The observed high energy properties of the PSR B1259–63 system as detected by GRO (G95, T96) are in good agreement with the calculated $L_\gamma$ and $\alpha$ of Fig. 10. We also notice that the spectral behavior in the hard X-ray energy range differ from that one in the 1-10 keV range, because of IC cooling affecting differently particles contributing to the soft and hard X-ray emission.

Models with either no IC cooling modification of $n(\gamma, t)$ (and therefore dominated by pure synchrotron cooling) or with unconstrained IC cooling for the whole orbit do not agree with observed data. An example of purely synchrotron cooling vs. combined synchrotron and IC cooling is given in Fig. 11, where the effect of IC cooling is clearly shown for the best model of Fig. 8. We find that cooling models characterized by pure synchrotron radiation cannot explain *simultaneously* the double hump of the X-ray emission and its spectral variability. On the other hand, models with combined synchrotron/IC cooling of the post-shock particle distribution function agree quite well with the observed data, but only for Be star outflow models with a thick equatorial disk misaligned with respect to the pulsar orbit. This is a general conclusion for the PSR B1259–63 system that applies to all the relevant parameter space of Table 3. Fig. 12 gives an example of the dependence of the results on the contrast between the polar and equatorial values of the outflow parameter. Fig. 13 shows how the results for the best misaligned model change as a function of the outflow index $n$. A value $n = 2.5$ is favored from a comparison with observed X-ray properties.

*Coplanar models*

Fig. 14 shows the predicted and observed X-ray properties for a coplanar model with $\Upsilon = 10^2$ (other coplanar models with different values of $\Upsilon$ share the same qualitative features). The figure shows both the synchrotron/IC cooling model (with characteristics similar to the misaligned model of Fig. 8, a choice that may appear unnatural in this case) and the purely synchrotron cooling model. It is clear that the purely synchrotron model does not agree with observations, and that the combined synchrotron/IC cooling might qualitatively explain the double hump of the X-ray luminosity light curve, even though the predicted luminosity is too high at *apastron* by a factor of $\sim 3$. This is a general conclusion for coplanar models of pulsar/outflow interaction.

Fig. 15 shows the predicted and observed X-ray properties for a coplanar model for different values of $\Upsilon$ assuming the most favorable combined synchrotron/IC cooling model. Notice how the predicted $L_x$ and $\alpha$ both become more and more dominated by pure synchrotron cooling for an increasing value of $\Upsilon$, reflecting a pulsar cavity inner boundary closer to the pulsar and further away from the Be star surface. We find that coplanar models of pulsar/outflow interaction are in disagreement with PSR B1259–63 data.

## 11. Discussion and conclusions

The wealth of radio X-ray and soft $\gamma$-ray data obtained for the PSR B1259–63 system at different orbital phases provide invaluable information to test the properties of the pulsar/outflow interaction. We find that:

(1) the PSR B1259–63 system turns out to be a *binary plerion* showing characteristics of *diffuse* and *compact* plerions near the apastron and periastron, respectively. The pulsar radiation and MHD wind pressure is able to avoid surface accretion or magnetospheric propeller-like processes throughout the whole period of high energy observations from Febrary 1992 through February 1994. A shock radius is established at relatively large distance from the pulsar, $r_s \sim 10^{12}$ cm (near periastron) corresponding to $\sim 10\%$ of the orbital distance. This radius corresponds to the distance of the 'apex' of the pulsar cavity in the plane of the pulsar orbit (which does not correspond to the line of sight direction). The pulsar wind pressure is able to 'break open' both the polar and equatorial parts of the Be star outflow (even at periastron), and a relatively dilute gaseous environment is produced along the line of sight for all orbital phases. There is no reason to invoke a non-spherical pulsar wind to a distance of $\sim 10^{12}$ cm. We also deduce that the apparent changes of pulsar spin period near periastron (M95) have an origin different from the suggested propeller effect (probably being the result of orbital parameter changes by tidal forces and induced by the covariance between orbital and pulsar parameters);

(2) the observed high-energy emission from the PSR B1259–63 system is in very good agreement with the predictions of a cooling model for the post-shock particle energy distribution function that assumes fast particle acceleration within a timescale smaller than the radiative timescales $\sim 10^2 - 10^3$ s. The index of the post-shock energy distribution function (before radiative cooling) is constrained to be constant throughout the PSR B1259–63 orbit and of value

$$p \simeq 2 \quad . \tag{82}$$

The best cooling model which explains simultaneously the X-ray intensity and spectral properties of the PSR B1259–63 system near periastron is given by a combination of synchrotron and IC cooling. We find that IC cooling is most effective near periastron where the 'screening' by the Be star outflow is at its minimum. Strong screening of the optical/IR flux relevant to IC cooling within the flow timescale of the pulsar cavity can occur in the equatorial disk. High-energy radiation is produced more effectively for cooled (by a factor of $\sim 10$ compared to the original post-shock energy) $e^{\pm}$-pairs of the pulsar wind being advected away in the inner parts of the pulsar cavity. The observable radiation in the ASCA-OSSE energy band is synchrotron emission, with the IC contribution calculated to be radiated above $\sim 10$ GeV near periastron. The range of particle energies ($[\gamma_1, \gamma_m]$) contributing to the ASCA-OSSE energy spectrum between 1 and 200 keV can be deduced as

$$10 \lesssim \frac{\gamma_m}{\gamma_1} \lesssim 100 \tag{83}$$

from the absence of detectable emission in the EGRET energy range near periastron (T96). Fig. 16 shows the calculated synchrotron emission for different values of the ratio $\gamma_m/\gamma_1$. As the pulsar recedes from periastron, the quantity $\gamma_m$ is expected to progressively increase for a synchrotron/IC power-law emission extending to energies $\gtrsim 10$ MeV. Note, however, that the overall shock-driven luminosity of the PSR B1259−63 system near apastron differs by $\sim 1$ order of magnitude compared to the periastron emission[8];

(3) the Be star outflow parameters are constrained to be $\Upsilon \sim 100$ and $n = 2.5$ at periastron. The region of the pulsar orbit near periastron (January 9, 1994) is most likely offset from the plane of the Be star equatorial outflow. We find that models assuming a coplanar geometry for the pulsar orbit and Be star equatorial outflow are not in agreement with X-ray data. In particular, the double hump of the X-ray luminosity light curve and the apastron/periastron contrast of intensity of about one order of magnitude cannot be reproduced by coplanar pulsar/outflow interaction models. Our best model is provided by a misaligned geometry, with a momentum flux contrast between polar and equatorial outflows (ratio $\Upsilon_2'/\Upsilon_1$) between $\sim 10$ and 100, and a relatively large angular width ($\Delta\theta_D \sim 50°$) of the equatorial disk at the pulsar orbital distance. The angle $i_D$ between the pulsar orbit and the centroid plane of the Be star equatorial outflow is constrained to be $i_D \gtrsim \Delta\theta_D/2 \sim 25°$. We note that our results are consistent with the independent analysis of Melatos *et al.* (1995) who studied the radio eclipse properties of PSR B1259−63 in detail. Radio and high-energy observations of the PSR B1259−63 system during the last four years are consistent with the value of the mass loss rate from the Be star companion being constant;

(4) time variable X-ray/soft $\gamma$-ray emission from the PSR B1259−63 system is established to originate from non-thermal particle acceleration processes most likely occurring at the pulsar wind termination shock. The efficiency $\xi$ of conversion of pulsar spindown power into visible radiation in the ASCA-OSSE energy range is $\xi \sim 4 - 5\%$, with a possible comparable contribution calculated to be in the MeV energy range (Paper II). This conversion efficiency is of the same order of magnitude as that one observed in the case of the Crab Nebula, suggesting a similar physical process of pulsar/nebula interaction and radiation. A series of constraints can be derived for both the pulsar wind parameters and for the details of the post-shock particle distribution function (Paper II). It is important to mention here that the acceleration mechanism timescale is constrained (for the first time for a relativistic object producing a MHD wind) to be less than $\sim 10^2 - 10^3$ seconds near periastron. We refer to Paper II for a discussion of the theoretical implications regarding the pulsar wind and non-thermal acceleration processes operating in the PSR B1259−63

---

[8]Future sensitive instruments in the 1-100 MeV energy range might detect the progressive hardening of the shock-driven spectrum of PSR B1259−63 as it recedes from periastron. The expected gamma-ray luminosity $L_\gamma \sim 10^{33}$ erg s$^{-1}$ near apastron is below the current detection capability of GRO instruments.

system.

Our results have general validity and they can be applied to a variety of astrophysical objects. The PSR B1259–63 system shows that X-ray emission in a binary system (not related to intrinsic stellar emission) can be efficiently produced by a non-thermal mechanism which is drastically different from accretion-driven or magnetospheric-driven emissions. Binary pulsars are ideal astrophysical systems in this context, because their shocked relativistic pulsar wind provides the crucial ingredient to produce high-energy emission of moderate luminosity. For a generic binary pulsar we expect the high-energy emission to be of a power-law form (unless the resulting plerion turns out to be ultra-compact) and strongly dependent on synchrotron and IC cooling. The $X/\gamma$-ray emission is predicted to be unpulsed and characterized by a small intrinsic column density, since nebular outflows which avoid gravitational 'trapping' near the compact star are expected to be optically thin along the line of sight in a way similar to PSR B1259–63.

We note that several low/intermediate luminosity X-ray and $\gamma$-ray sources in our Galaxy might contain a pulsar similar to PSR B1259–63 orbiting around a low-mass or high-mass companion (e.g, Tavani, 1995). PSR B1259–63 is orbiting around its Be star companion in a very eccentric and long orbital period orbit, with its radio emission detectable during most of its revolutions, except for the 40-day eclipses near periastron. It is easy to imagine similar binary pulsars (with low-mass or high-mass companions) of smaller orbital periods and therefore likely to be eclipsed for most if not all of their orbit. A selection effect against the radio discovery of pulsars in compact binaries has been considered in general (Tavani, 1991). We note that a more focused search for these systems can be attempted today in the $X/\gamma$-ray energy range after the results obtained for PSR B1259-63. Interesting sources may include a class of time variable unidentified EGRET sources near the Galactic plane (Tavani *et al.* 1996b), dim X-ray sources in globular clusters (e.g., Hasinger *et al.* 1994, Grindlay 1995), and OB-associations with anomalously large X-ray emission (e.g., Motch & Pakull, 1996). All these systems show significant excess of their high-energy emission over the expected level of stellar emission. From the results obtained for the PSR B1259–63 system, we expect that temporal variations of the intensity and spectrum of Galactic and $X/\gamma$-ray unidentified sources can provide evidence for underlying non-thermal processes powered by 'hidden' pulsars.

All current data on the PSR B1259–63 system favor an outflow model from its massive companion characterized by a constant mass loss rate. However, we know that this is not the case in general, and that there are Be systems with rapidly spinning neutron stars (such as A0538-66) that display large variations of their mass loss rate. Be star systems with pulsars can therefore show different 'outflow states', and corresponding 'low' and 'high' level of high-energy emission. We believe that, despite the large variety of possible systems, hidden pulsars can be searched today in their 'low' states because of the distinctive features of the high-energy emission as clearly shown by the PSR B1259–63 system. Future monitoring of the PSR B1259–63 system will provide valuable information on the long timescale variability of the mass loss rate of SS 2883.

**Acknowledgements**

We would like to thank our colleagues of the radio, ASCA and GRO observing teams for their assistance in the multiwavelength campaign at the periastron passage of PSR B1259-63. We are especially grateful to R.N. Manchester and S. Johnston for exchanges and discussion of radio data, and to F. Nagase, V. Kaspi, E. Grove and M. Hirayama for their collaborative work on the high-energy observing campaign near periastron. M. Hirayama kindly provided Figures 1-3. M. Tavani's research was supported by NASA grants NAG 5-2593 and NAG 5-2730; that of J. Arons was supported by NASA grant NAGW-2413 and by NSF grant AST-9115093. Part of this work was supported under the auspices of the U.S. Department of Energy at the Lawrence Livermore National Laboratory under contract W-7405-Eng-48.

# References

Arons J., 1983, Nature 302 301.

Arons, J., 1992, in The Magnetospheric Structure and Emission Mechanism of Radio Pulsars, IAU Colloq. no. 128, eds. J.A. Gil & J.M. Rankin (Zielona Góra: Pedagogical Univ. Press) p. 56.

Arons, J., 1996, in Pulsars: Problems and Progress, IAU Colloq. no. 160, eds. M. Bailes, S. Johnston & M. Walker (San Francisco: Astronomical Society of the Pacific), in press.

Arons J., Tavani M., 1993, ApJ, 403, 249 (AT93).

Baranov, V.B., Krasnobaev, K.V. & Kulikovskii, A.G., 1971, Sov. Phys. Doklady, 15, 791.

Bell, J.F., Bailes, M., Manchester, R.N., Weisberg, J.M. & Lyne, A.G., 1995, ApJ, 440 L81.

Bjorkman, J.E. & Cassinelli, J.P., 1993, ApJ, 403, 429.

Blumenthal G., Gould R., 1970, Rev. Mod. Phys, 42, 237.

Brookshaw, L. & Tavani, M., 1995, in Millisecond Pulsars: a Decade of Surprise, eds. A. Fruchter, M. Tavani, D.C. Backer (San Francisco: Astronomical Society of the Pacific), p. 244.

Burnard, D.J., Lea, S.M., and Arons, J. 1983, ApJ, 266, 175.

Campana, S., Stella, L., Mereghetti, S. & Colpi, M., 1995, A&A, 297, 385.

Cassinelli, J.P., Cohen, D.H., MacFarlane, J.J., Sanders, W.T. & Welsh, B.Y., 1994, ApJ, 421, 705.

Clear, J., *et al.*, 1987, A&A, 174, 85.

Coe, M.J., *et al.*, 1993, in AIP Conference Proceedings no. 280, eds. M. Friedlander, N. Gehrels, D.J. Macomb, p. 360.

Cominsky L., Roberts, M., Johnston, S., 1994, ApJ, 427, 978.

Cordes, J.M., 1996, in Pulsars: Problems and Progress, IAU Colloq. no. 160, eds. M. Bailes, S. Johnston & M. Walker (San Francisco: Astronomical Society of the Pacific), in press.

Cordes, J.M., Romani, R.W. & Lundgren, S.C., 1993, Nature, 362, 133.

Emmering, & Chevalier, 1987, ApJ, 321, 334.

Felten J.E. & Morrison, P., 1966, ApJ, 146, 686.

Gallant, Y.A., & Arons, J. 1994, ApJ, 435, 230 (GA94).

Ghosh, P, 1995, ApJ, 453, 411.

Grindlay, J., 1995, in Millisecond Pulsars: a Decade of Surprise, eds. A. Fruchter, M. Tavani, D.C. Backer (San Francisco: Astronomical Society of the Pacific), p. 57.

Grove, J.E., Tavani, M., Purcell, W.R., Johnson, W.N., Kurfess, J.D., Strickman, M.S. & Arons, J., 1995, ApJ, 447, L112 (G95).

Greiner, J., Tavani, M., Belloni, T., 1995, ApJ, 441, L43.

Hasinger, G., Johnston, H. & Verbunt, F., 1994, A&A, 288, 466.

Hester, J.J., *et al.*, 1995, ApJ, 448, 240.

Hills, J.G., 1983, ApJ, 267, 322.

Hirayama, M., Nagase, F., Tavani, M., Kaspi, V.M., Kawai, N. & Arons, J., 1996, PASP, in press (H96).

Hoshino, M., Arons, J., Gallant, Y.A. & Langdon, A.B., 1992, ApJ, 390 454 (HAGL92).

Illarionov A.F. & Sunyaev, R.A., 1975, A&A, 39, 185.

Johnston S., Manchester R. N., Lyne A. G., Bailes M., Kaspi V. M., Qiao G., D'Amico N., 1992, ApJ, 387, L37 (J92).

Johnston S., Manchester R. N., Lyne A. G., Nicastro, L. and Spyromilio, J., 1994, MNRAS, 268, 430 (J94).

Johnston, S., *et al.*, 1996, MNRAS, 279, 1026 (J96).

Jones, T.W. & Hardee, P.E., 1979, ApJ, 228 268.

Kaspi, V., Tavani, M., Nagase, F., Hoshino, M., Aoki, T., Kawai, N. & Arons, J., 1995, ApJ, 453, 424 (KTN95).

Kardashev, N., 1962, Sov. Astron.-AJ, 6 , 317.

Kennel C. F., Coroniti F. V., 1984, ApJ, 283, 694 (KC84).

King A., 1993, ApJ, 405, 727.

King A., & Cominksy, L., 1994, ApJ, 435, 411 (KC94)

Kochanek C., 1993, ApJ, 406, 638.

Kulkarni, S. & Hester, 1988, Nature, 335, 801.

Lipunov, V.M., Nazin, S.N., Osminkin, E.Yu. & Prokhorov, M.E., 1994, A&A, 282, 61.

Makino F. & Aoki, T., 1994, personal communication.

Manchester, R.N., 1994, in Pulsars, eds. G. Srinivasan etal (Bangalore: Indian Academy of Sciences), in press.

Manchester, R.N., Johnston, S., Lyne, A.G., D'Amico, N., Bailes, M. & Nicastro, L., 1995, ApJ, 445, L137 (M95).

Melatos, A, Johnston, S. & Melrose, D.B., 1995, MNRAS, 275, 381.

Michel, 1969, ApJ, 158, 727.

Motch, C. & Pakull, M., 1996, IAC Circ. no. 6285.

Nolan, P.L., *et al.*, 1993, ApJ, 409, 697.

Ray, P.S., *et al.*, 1993, in Compton Gamma-Ray Observatory Symposium, eds. M. Friedlander, N. Gehrels & D.J. Macomb (New York: AIP Conf. Proc. no. 280), p. 249.

Skinner *et al.*, 1980, Ap.J. 240, 619.

Tavani, M., 1991, ApJ, 379, L69.

Tavani, M., 1995, in *The Gamma-Ray Sky with COMPTON GRO and SIGMA*, eds. M. Signore, P. Salati & G. Vedrenne (Dordrecht: Kluwer Academics), p. 181.

Tavani, M., Arons, J., Kaspi, V., 1994, ApJ, 433, L37 (TAK94).

Tavani, M., Grove, J.E., Purcell, W., Hermsen, W., Kuiper, L., Kaaret, P., Ford, E., Wilson, R.B., Finger, M., Harmon, B.A., Zhang, S.N., Mattox, J., Thompson, D & Arons, J., 1996a, A&AS, in press (T96).

Tavani, M., Mukherdjee, R., Mattox, J., *et al.*, 1996b, to be submitted to ApJ.

Tavani, M. & Arons, J., 1996, to be submitted to ApJ (Paper II).

Taylor, J.H., Manchester, R.N. & Lyne, A.G., 1993, ApJS, 88, 529.

Thompson, D., et al., 1994, ApJ, 436, 229 (Th94).

Waters L. B. F. M., 1986, A&A, 162, 121.

Waters L. B. F. M., Taylor A. R., van den Heuvel E. P. J., Habets G. M. H. J., Persi P., 1988, A&A, 198, 200 (W88).

Waters L. B. F. M., de Martino, D., Habets, G.M.H.J. & Taylor, A.R., 1989, A&A, 223, 207.

**Table 1: Pulsar and Orbital Parameters of the PSR B1259−63 System[a]**

| | |
|---|---|
| Right Ascension, $\alpha$ (J2000) | 13h 02m 47.68s |
| Declination, $\delta$ (J2000) | −63° 50' 08".6 |
| Dispersion Measure, DM | 146.72(3) pc cm$^{-3}$ |
| Pulsar Period, $P$ | 47.762053542 (8) ms |
| Period Derivative, $\dot{P}$ | 2.27579 (16) $\times 10^{-15}$ |
| Period Epoch | MJD 48053.440 |
| Spindown Age, $\tau$ | $3 \times 10^5$ yr |
| Magnetic Field, $B$ | $3 \times 10^{11}$ G |
| Spindown Luminosity, $L_p$ | $8 \times 10^{35}$ erg s$^{-1}$ |
| Orbital Period, $P_{\rm b}$ | 1236.772359(5) day |
| Projected semi-major axis, $a_{\rm p} \sin i$ | 1296.580 (2) lt s |
| Longitude of periastron, $\omega$ | 138.6782 (2)° |
| Eccentricity, $e$ | 0.869931 (1) |
| Periastron Epoch, $T_0$ | MJD 48124.3448 (1) |

[a] From Manchester *et al.* (1995).

**Table 2: Comparison between the PSR B1259−63 and A0538-66 systems**

|  | PSR B1259-63 | A0538-66 high state | A0538-66 low state |
|---|---|---|---|
| Pulsar spin period (s) | 0.047 | 0.069 | |
| Spindown luminosity (erg s$^{-1}$) | $8 \cdot 10^{35}$ | ? | |
| X-ray luminosity (erg s$^{-1}$) | $\sim 10^{34}$ | $\sim 10^{38}$ | $\lesssim 10^{35}$ |
| X-ray pulsations | no | yes | ? |
| Column density $N_H$ (in $cm^{-2}$) | $5 \cdot 10^{21}$ | $2 \cdot 10^{23}$ | ? |
| Photon spectral index (1-10 keV) | 1.6-1.9 | 0.6-1.6 | ? |

Data from KTN95 and H96 for PSR B1259−63 and from Skinner *et al.* (1981) for A0538-66.

**Table 3: Model parameters and comparison with ASCA/GRO data[⋆]**

| Quantity | Range | Comparison with data |
|---|---|---|
| *Coplanar models* | | |
| Outflow parameter $\Upsilon$ | 10 | no agreement for $L_x$ and $\alpha_X$ |
| (for $2 \leq n \leq 4$) | 100 | no agreement for $L_x$ and $\alpha_X$ |
| | $10^3$ | strong disagreement for $L_x$ and $\alpha_X$ |
| Initial outflow velocity $v_0$ | $\gtrsim 10^7 \,\mathrm{cm\,s^{-1}}$ | marginal agreement for $L_x$ and $\alpha_X$ |
| | $10^6 \,\mathrm{cm\,s^{-1}}$ | no agreement for $L_x$ and $\alpha_X$ |
| IC cooling suppressed | | strong disagreement for $L_x$ and $\alpha_X$ |
| IC cooling ($|\phi| < \sigma_{ic}$) | $\sigma_{ic} = 50°$ | marginal agreement with $L_x$ and $\alpha_X$ (low $\Upsilon$) |
| | $\sigma_{ic} = 180°$ | strong disagreement for $\alpha_X$ |
| *Misaligned models* | | |
| Outflow parameters $\Upsilon_1, \Upsilon_2$ | $\Upsilon_1 = 10^2, \Upsilon_2' \sim 10^3$ | good agreement |
| | $\Upsilon_1 = 10^2, \Upsilon_2' \sim 10^4$ | marginal disagreement for $\alpha_X$ |
| Initial outflow velocity $v_0$ | $\gtrsim 10^7 \,\mathrm{cm\,s^{-1}}$ | good agreement for $L_x$ and $\alpha_X$ |
| | $\sim 10^6 \,\mathrm{cm\,s^{-1}}$ | no agreement for $L_x$ and $\alpha_X$ |
| Disk angular size at $r_s$ | $\sigma_d(r_s) = 30°$ | good agreement |
| First interaction point $\theta_1$ | $\theta_1 = 120°$ | good agreement |
| Outflow index $n$ | 2.5 | good agreement |
| | 3 | poor agreement for $L_x$ and $\alpha_X$ |
| | 4 | no agreement for $L_x$ and $\alpha_X$ |
| IC cooling suppressed | | strong disagreement for $L_x$ and $\alpha_X$ |
| IC cooling ($|\phi| < \sigma_{ic}$) | $\sigma_{ic} = 50°$ | good agreement |
| | $\sigma_{ic} \gtrsim 60°$ | no agreement for $L_x$ and $\alpha_X$ |

($\star$) Pulsar wind parameters have been fixed to $\gamma_1 = 10^6$ and $\sigma = 0.02$ (see Paper II). The comparison with data for each model is meant to be done for that particular value of the model parameter taking all other parameters to be equal to those specified first in the table.

Fig. 1.— GINGA, ROSAT and ASCA observations of the PSR B1259–63 system (from Hirayama *et al.*, 1996). The ROSAT observations near apastron were carried out in Feb. 25-26 1992 (obs1), Aug. 30-Sept. 4 1992 (obs2), and Feb. 7-16, 1993 (obs3) (CRJ94, Greiner *et al.*, 1995). ASCA observations near periastron were carried out in Dec. 28 1993 (obs1), Jan. 10 1994 (obs2), Jan. 26 1994 (obs3), and Feb. 28 1994 (obs4). *Compton* GRO observations were carried out during the period Jan. 3-23 1994, i.e., between ASCA obs1 and obs3 (T95).

Fig. 2.— Summary of ROSAT and ASCA X-ray observations of the PSR B1259–63 system (from Hirayama *et al.* 1996; KTN95; CRJ94; Greiner *et al.* 1995). The figure shows the time variation of the X-ray luminosity in the ROSAT (0.1-2.5 keV) and ASCA (1-10 keV) energy bands, as a function of orbital phase. See Fig. 1 for a definition of ROSAT and ASCA observations. Also reported is the GINGA upper limit of Sept. 5 1991 (Makino & Aoki, 1994).

Fig. 3.— Time behavior of the column density $N_H$ and of the photon spectral index $\alpha$ of X-ray emission (1-10 keV) for a power-law fit to ASCA data of the PSR B1259–63 system near periastron (from Hirayama *et al.*, 1996; see also KTN95). The countours give the $1\sigma$, $2\sigma$, and $3\sigma$ levels for a spectral fit giving simultaneously $N_H$ and $\alpha$. The modified Julian dates marking the different observations refer to the ASCA observations (see Fig. 1).

Fig. 4.— Combined ASCA and CGRO photon spectra of the PSR B1259–63 system near periastron (Tavani, 1996). The ASCA spectrum (covering the 1-10 keV eneregy band) is for the post-periastron January 26 1994 observation (KTN95). The ASCA spectrum at periastron is slightly softer than the spectrum shown here and reduced by a factor of $\sim 2$. The pre-periastron ASCA observation of December 28, 1993 showed intensity and spectral characteristics very similar to the January 26, 1994 observation. The CGRO multi-instrument observations cover the time period January 3-23, 1994. The OSSE instrument (of energy range between 50 keV and a few MeV) is the only CGRO detector capable of detecting emission from the PSR B1259–63 system at a few mCrab level up to $\sim 200$ keV (G95). COMPTEL and EGRET upper limits to the emission are also reported (T96).

Fig. 5.— Coplanar model of the PSR B1259–63 binary for $\Upsilon = 10$. Shock radius, gravitational capture radius, ions' Larmor radius, compressed local magnetic field at the shock radius and radiation timescales calculated as a function of orbital phase ($\theta = \phi + \pi$, where $\phi$ is the true anomaly and $\theta = 180°$ at periastron). The model presented in this figure assume $m_2 = 10\,M_\odot$, $n = 2.5$, $v_0 = 10^7\,\mathrm{cm\,s^{-1}}$, $\gamma_1 = 10^6$ and $\sigma = 0.02$. *Left plot:* Values of the magnetic field $B_2(r_s)$ (in Gauss) and of the ratio of adiabatic and acceleration timescales $\tau_1/\tau_{\mathrm{acc}}$. *Central plot:* calculated values of the orbital distance $d$, shock radius $r_s$, ions' Larmor radius $R_L$, gravitational capture radius $R_{grav}$ (all given in cm). *Right plot:* calculated values (in sec) of the adiabatic timescale $\tau_{\mathrm{ad}}$, ions' acceleration timescale $\tau_{\mathrm{acc}}$ (from Eq. 24), synchrotron timescale $\tau_s$ (from Eq. 25), and relativistic inverse Compton scattering timescale $\tau_{icr}$ (from Eq. 31).

Fig. 6.— Coplanar model of the PSR B1259–63 binary. Same as Fig. 5 for $\Upsilon = 10^2$.

Fig. 7.— Coplanar model of the PSR B1259–63 binary. Same as Fig. 5 for $\Upsilon = 10^3$.

Fig. 8.— Misaligned model of the PSR B1259–63 system for tilted pulsar orbital plane and Be star mass equatorial outflow. Shock radius $r_s$, ions' Larmor radius $R_L$, compressed local magnetic field at the shock radius $B_2(r_s)$ and radiation timescales calculated as a function of orbital phase ($\theta = 180$ at periastron), for $\Upsilon_1 = 10^2$, $\Upsilon'_2 = 7 \cdot 10^2$, $\sigma_d = 35°$, $\sigma_{ic} = 50°$, $m_2 = 10\,M_\odot$, $n = 2.5$, $\gamma_1 = 10^6$ and $\sigma = 0.02$. *Left plot:* Values of the magnetic field $B_2(r_s)$ (in Gauss) and of the ratio of adiabatic and acceleration timescales $\tau_{\rm ad}/\tau_{\rm acc}$. *Central plot:* calculated values of the orbital distance $d$, ions' Larmor radius $R_L$, and shock radius $r_s$ (in cm). *Right plot:* calculated values (in sec) of the adiabatic timescale $\tau_{\rm ad}$, ions' acceleration timescale $\tau_{\rm acc}$ (from Eq. 24), synchrotron timescale $\tau_s$ (from Eq. 25), and relativistic IC scattering timescale $\tau_{icr}$ (from Eq. 31). and IC scattering timescale $\tau_{ic}$ for post-shock adiabatically cooled particles calculated in an intermediate regime for an assumed value of $\alpha_{ic} = 10$.

Fig. 9.— Observed and calculated values of the X-ray luminosity and photon spectral index $\alpha$ in the energy band 2-10 keV as a function of orbital phase for the misaligned model of Fig. 8. Periastron is at $\theta = 180$. Both synchrotron and IC cooling processes are included with $\sigma_{ic} = 50°$. *(Solid curves:)* results of the calculation for $\alpha_{ic} = 10$; *(dashed curves:)* results of the calculation for $\alpha_{ic} = 5$. *(Upper plot:)* ASCA X-ray luminosity (1$\sigma$ error bars) and calculated shock luminosity in the energy band 1-10 keV. The overall normalization of the calculated curves reflects a conversion efficiency of pulsar spindown into X-rays of $\sim 1\%$. *(Lower plot:)* spectral photon index $\alpha$ as determined by ASCA and calculated for the misaligned model of Fig. 8 for $\varepsilon = 5$ keV.

Fig. 10.— Same as Fig. 9 for the calculated intensity and spectral properties of the hard X-ray emission with $\varepsilon = 100$ keV. The calculated luminosity and photon spectral index are for the misaligned model of Fig. 8 with $\alpha_{ic} = 100$ reflecting the inefficient IC cooling in the relativistic regime for the particles with energies contributing to synchrotron emission in the hard X-ray range. The thick (light) arrow marks the beginning (end) of the continuous *Compton* GRO observation of the PSR B1259–63 system (G95, T96). The average intensity and spectral index in the 50-200 keV range detected by OSSE are $L_\gamma \sim 3 \cdot 10^{34}$ erg s$^{-1}$, and $\alpha = 0.8 \pm 0.6$ (G95).

Fig. 11.— Comparison between calculated and observed properties of X-ray emission for different assumption on the main cooling mechanism near periastron. Same as Fig. 9 for quantities calculated for $\varepsilon = 5$ keV and $\alpha_{ic} = 5$. *(Solid curves:)* model with both IC and synchrotron cooling for $\sigma_{ic} = 50°$; *(dashed curves:)* model with only synchrotron cooling.

Fig. 12.— Comparison between calculated and observed properties of X-ray emission for different values of the outflow parameter $\Upsilon'_2$ for a misaligned model. All other parameters are the same as the model of Fig. 8 ($\Upsilon_1 = 100$), and the plotted quantities are the same as those in Fig.9 with $\varepsilon = 5$ keV, $\alpha_{ic} = 5$, and $\sigma_{ic} = 50°$. *(Solid curves:)* model with $\Upsilon'_2 = 7 \cdot 10^2$; *(short-dashed curves:)* model with $\Upsilon'_2 = 7 \cdot 10^3$; *(long-dashed curves:)* model with $\Upsilon'_2 = 7 \cdot 10^4$.

Fig. 13.— Comparison between calculated and observed properties of X-ray emission for different values of the outflow index $n$. All other parameters are the same as the model of Fig. 8 and the plotted quantities are the same as those in Fig.9 with $\varepsilon = 5$ keV and $\alpha_{ic} = 10$ keV. *(Solid curves:)* model with $n = 2$; *(short-dashed curves:)* model with $n = 3$; *(long-dashed curves:)* model with $n = 4$.

Fig. 14.— Coplanar model with $\Upsilon_1 = 10^2$ and $\Upsilon_2' = 0$. Observed and calculated values of the X-ray luminosity and photon spectral index $\alpha$ (for $\varepsilon = 5$ keV) as a function of orbital phase. (periastron is at $\theta = 180$). Other parameters are as in Fig. 8.

*(Solid curves:)* results of a calculation with both IC and synchrotron cooling included (assuming no substantial absorption of the optical flux from the companion surface near periastron and the same absorption behavior of the optical flux as in Fig. 8); *(dashed curves:)* results for a calculation with only synchrotron cooling included (assuming strong absorption of the optical flux from the companion surface and no IC effect of the IR flux within the flow timescale of the pulsar cavity).

*(Upper plot:)* calculated and observed X-ray luminosity (1-10 keV). *(Lower plot:)* observed and calculated spectral photon index $\alpha$ calculated for $\varepsilon = 5$ keV.

Fig. 15.— Coplanar model for different values of $\Upsilon_1$. Observed and calculated values of the X-ray luminosity and photon spectral index $\alpha$ (in the energy band 2-10 keV) as a function of orbital phase. (periastron is at $\theta = 180$). Parameters other than $\Upsilon_1$ are as in Fig. 8. The results of the calculation presented here assume no substantial absorption of the optical flux from the companion surface and efficient IC cooling as given for Fig. 8. *(Solid curves:)* results for $\Upsilon_1 = 10^2$; *(small-dashed curves:)* results for $\Upsilon_1 = 10^3$; *(large-dashed curves:)* results for $\Upsilon_1 = 10^4$. *(Upper plot:)* observed and calculated X-ray luminosity (1-10 keV). *(Lower plot:)* observed and calculated spectral photon index $\alpha$ calculated for $\varepsilon = 5$ keV.

Fig. 16.— Calculated synchrotron spectra compared to the observed high-energy emission from the PSR B1259–63 system. Data are from ASCA (*starred points*; KTN95), OSSE (*filled triangles*; G95); COMPTEL (*open triangles*, T96); EGRET (*open squares*, T96). The superimposed curves give the calculated synchrotron emission spectrum for $\epsilon_c =1$ keV, average spectral index of the post-shock particle distribution function $p = 2.5$, and ratio of maximum to initial post-shock particle energy $\gamma_m/\gamma_1 = 10$ (*solid curve*), $\gamma_m/\gamma_1 = 100$ (*short dashed curve*), $\gamma_m/\gamma_1 = 10^3$ (*long dashed curve*). An effective column density $N_H = 6 \cdot 10^{21}$ cm$^{-2}$ is assumed.

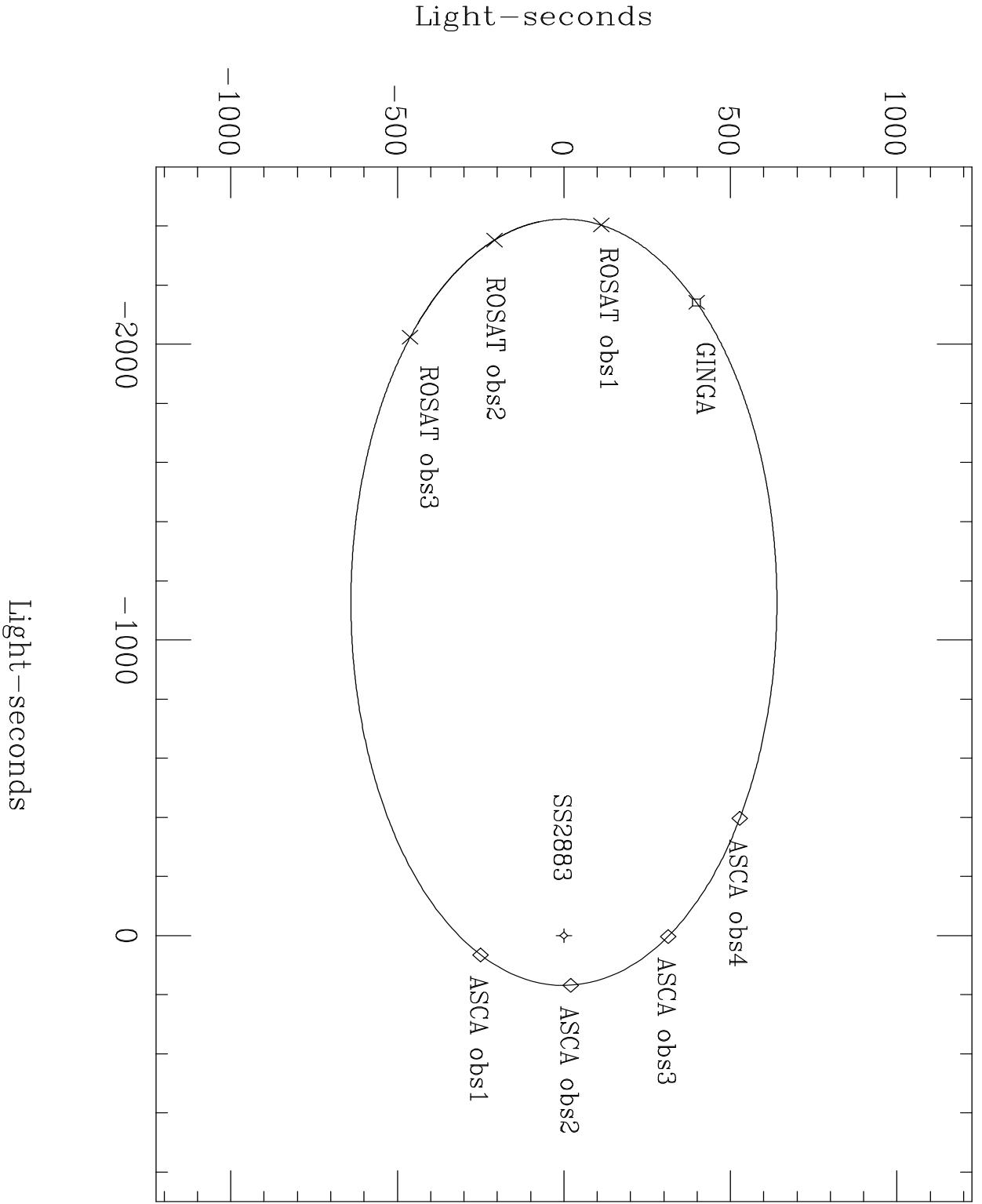

Fig. 1

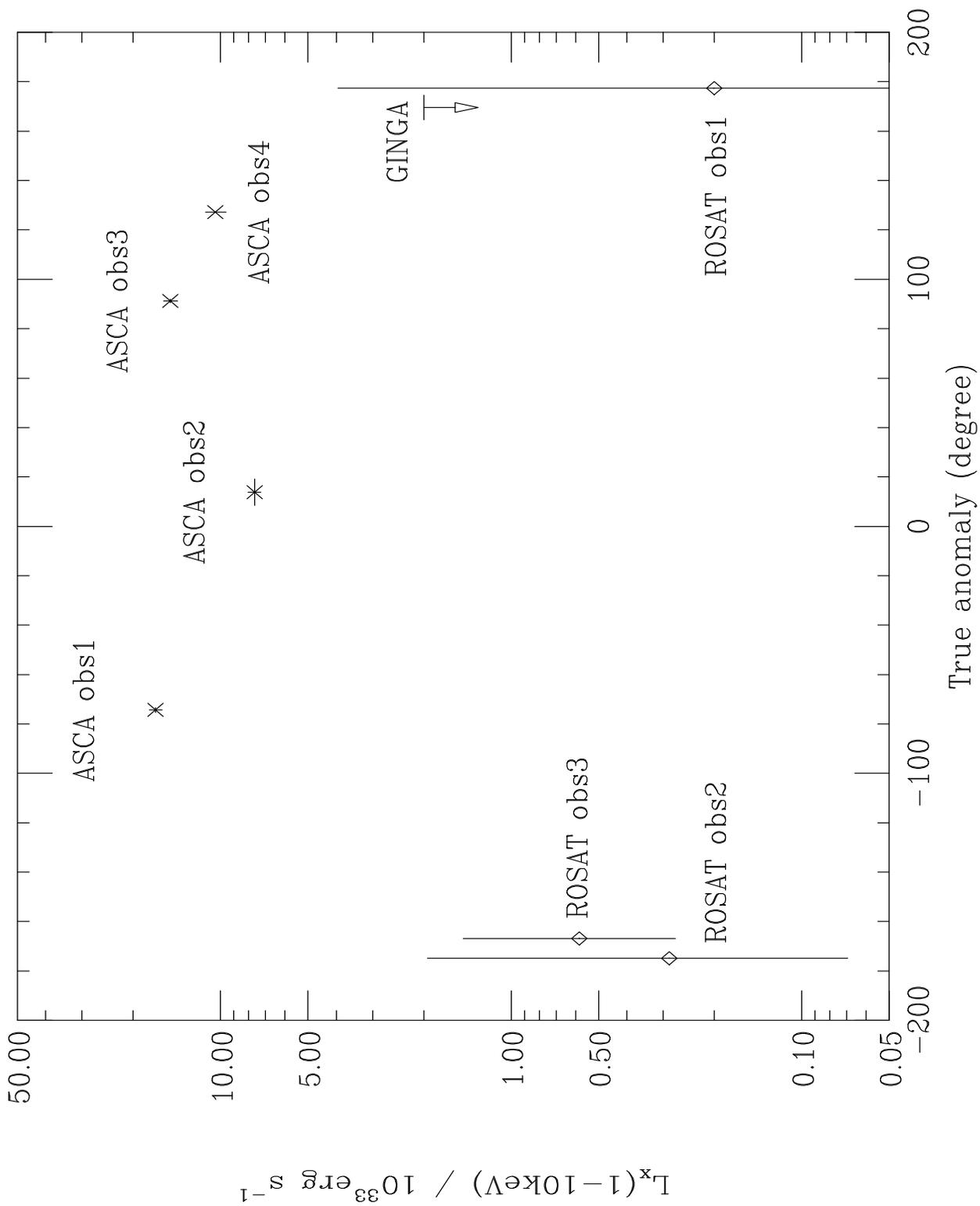

Fig. 2

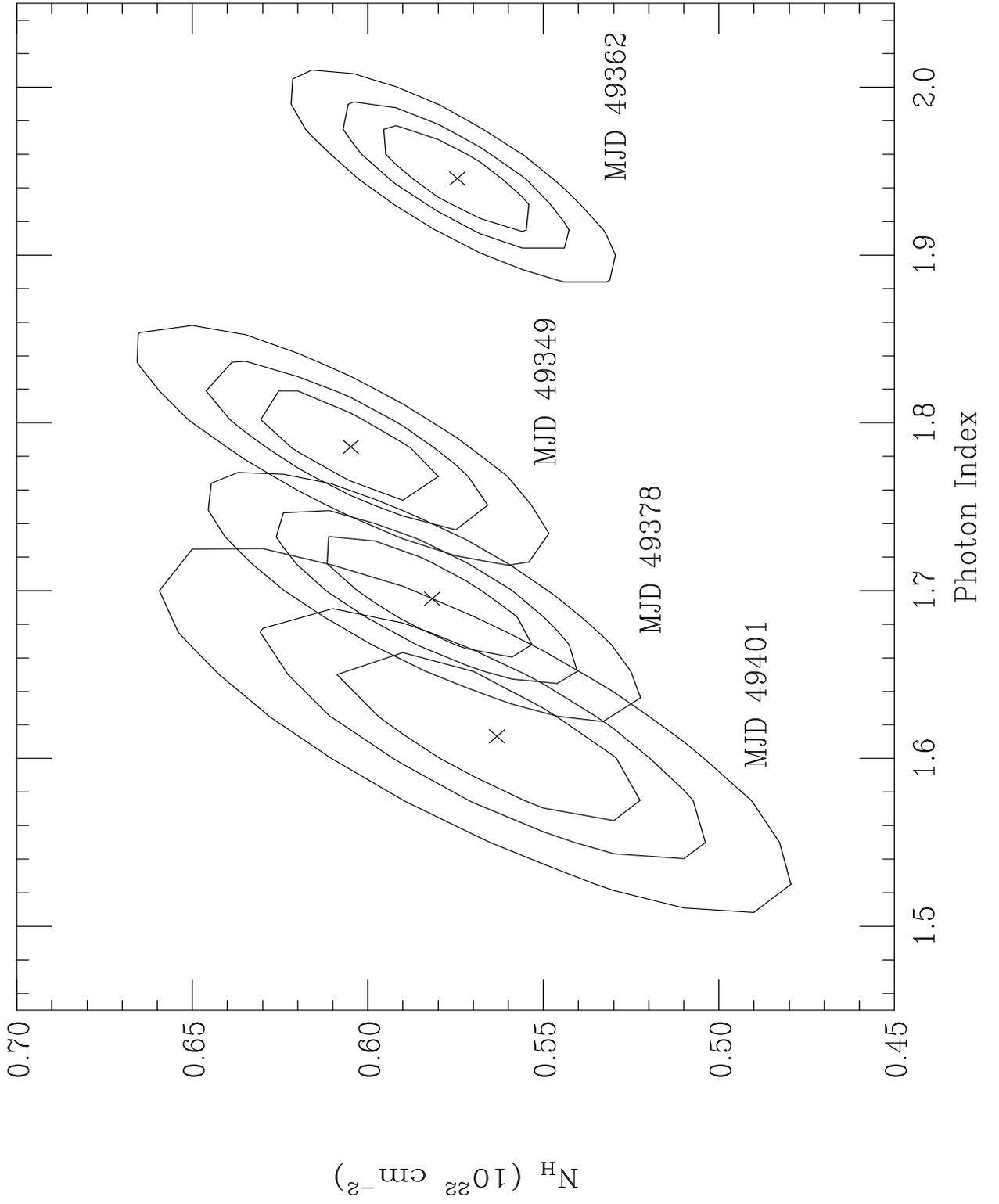

Fig. 3

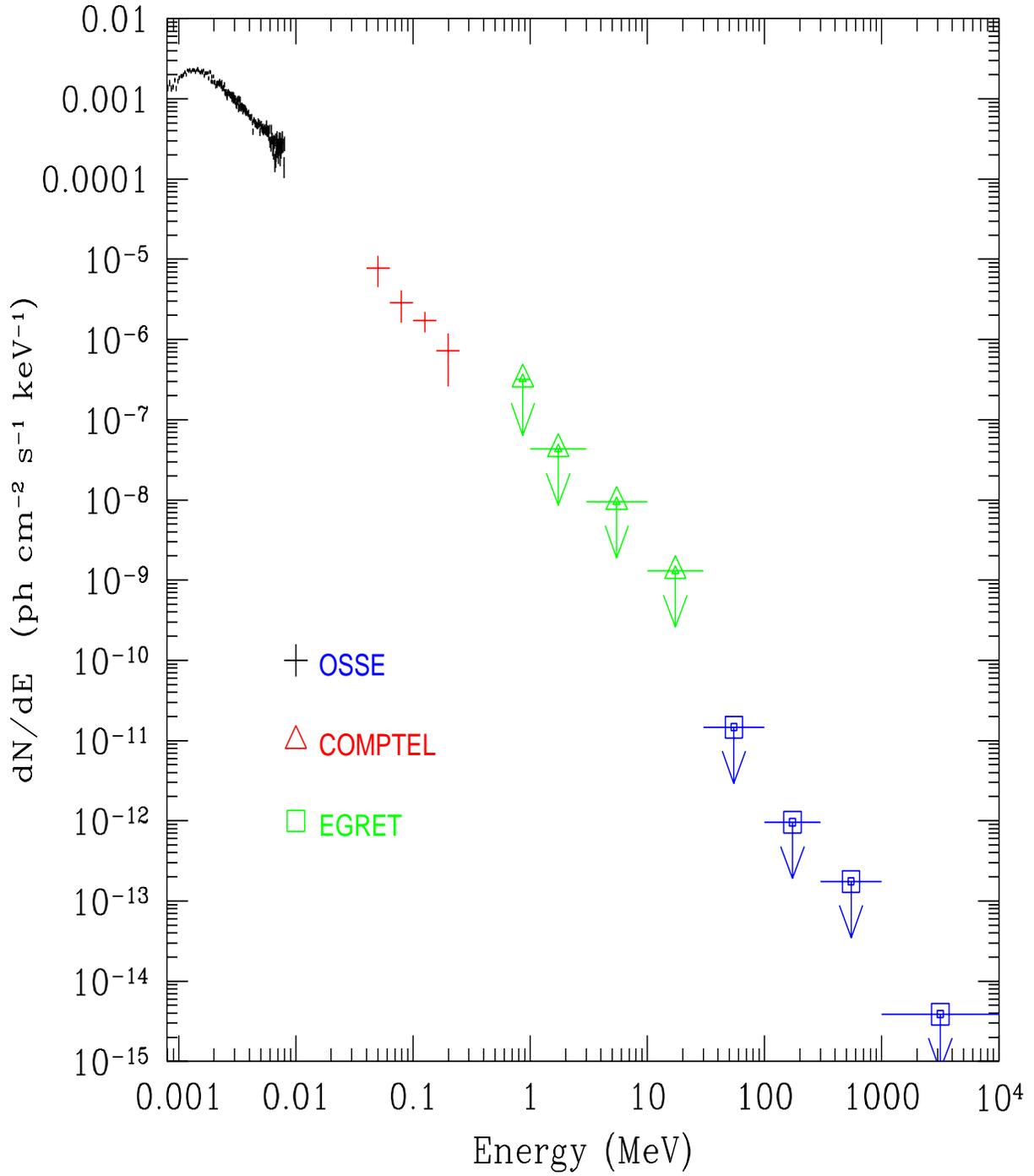

Fig. 4

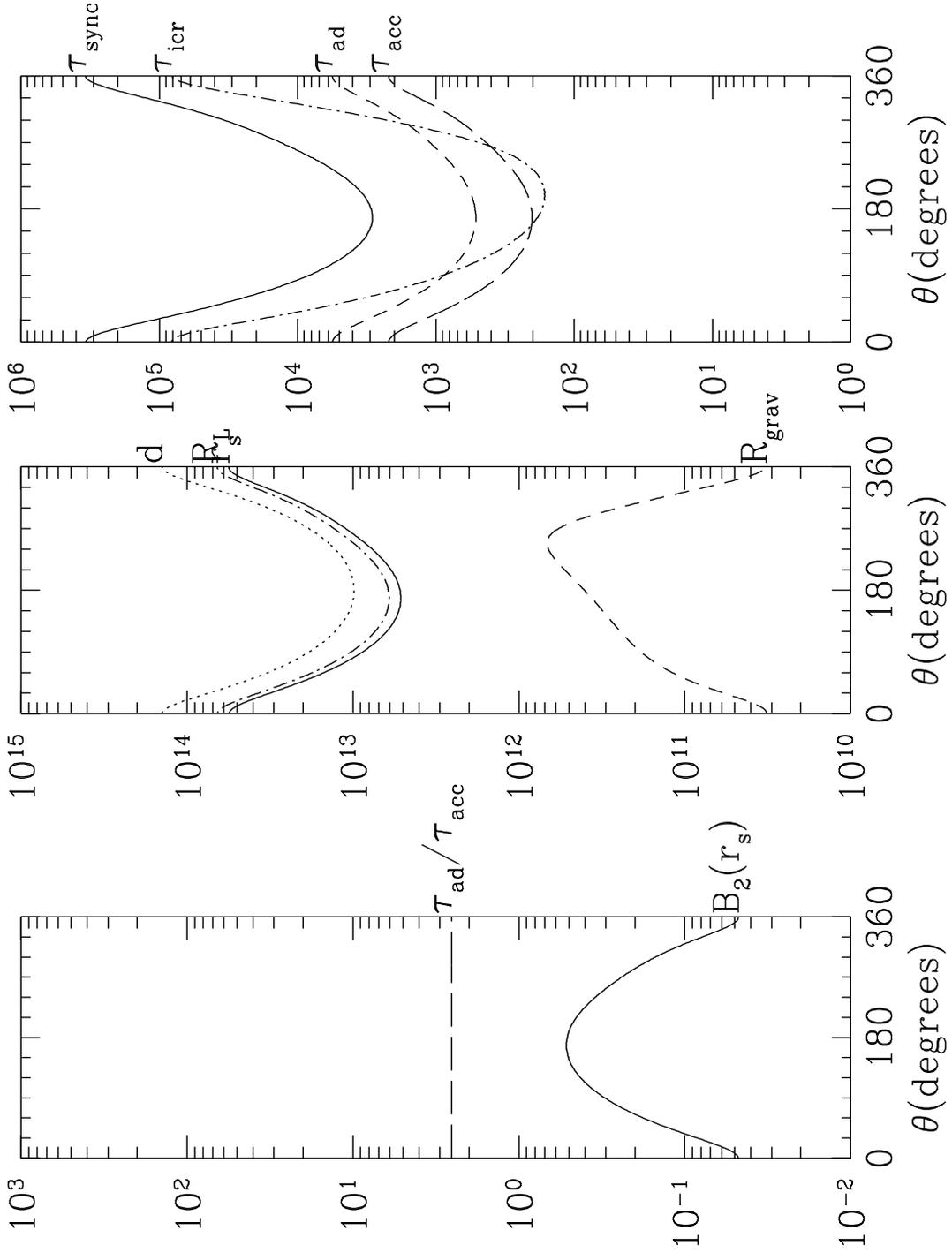

Fig. 5

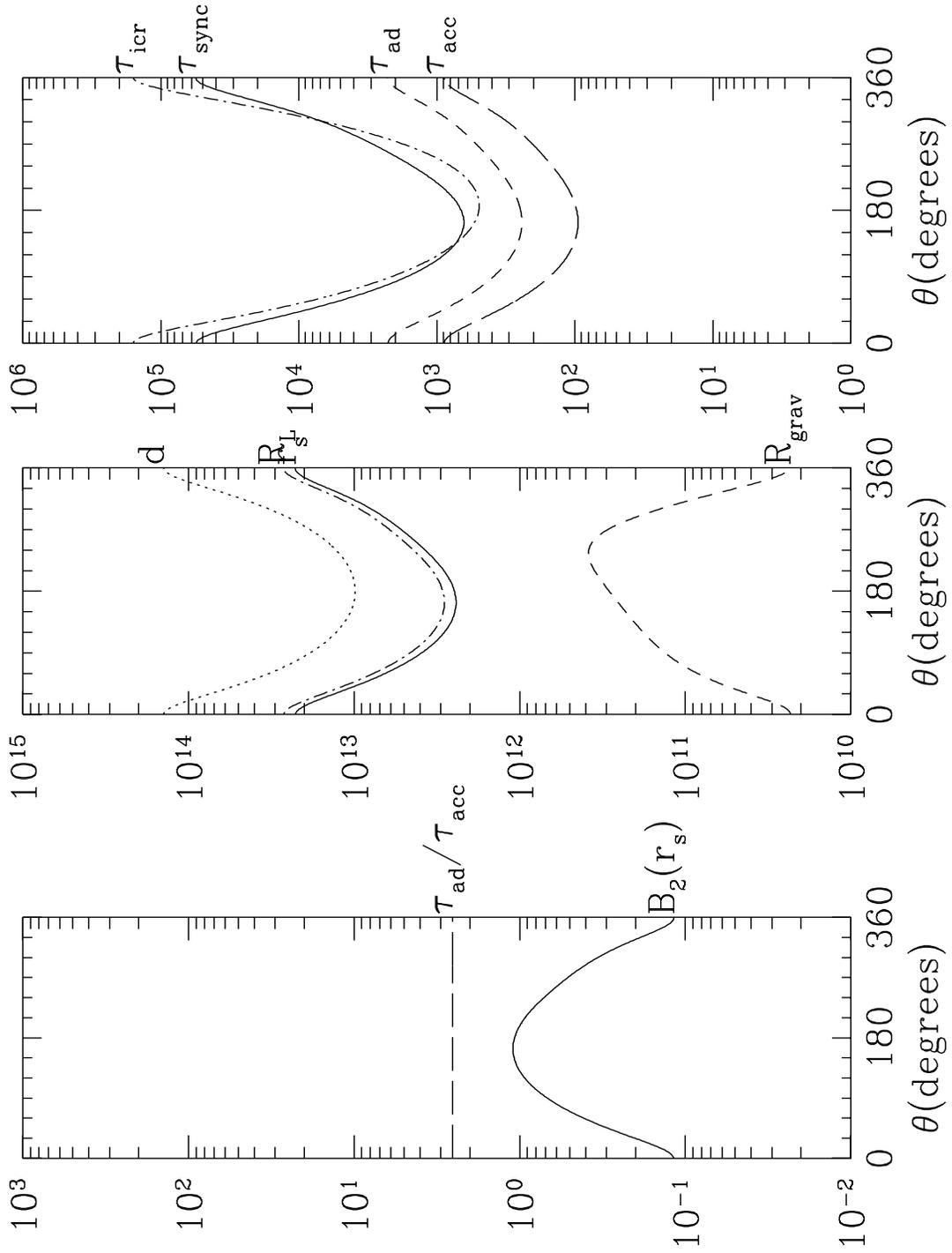

Fig. 6

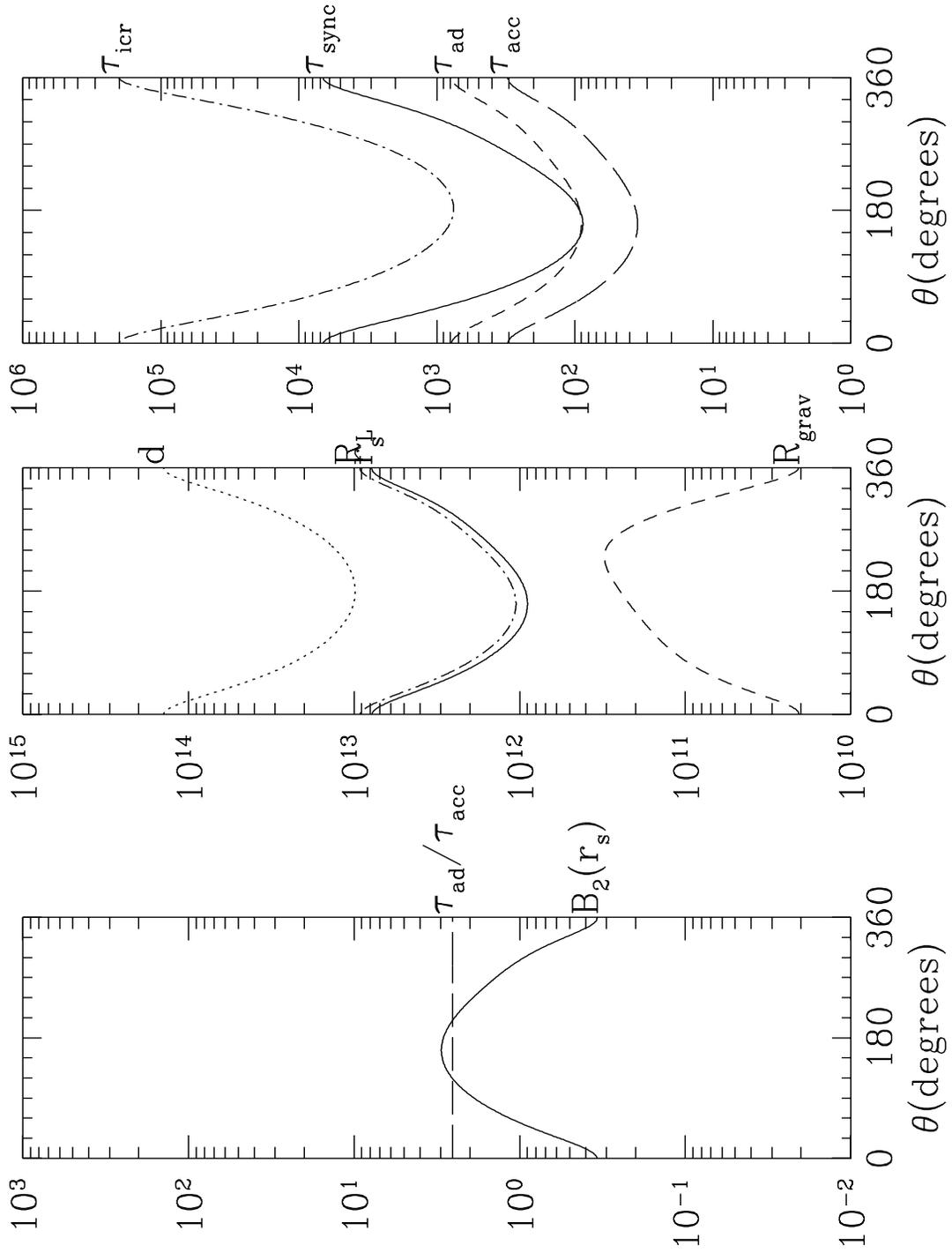

Fig. 7

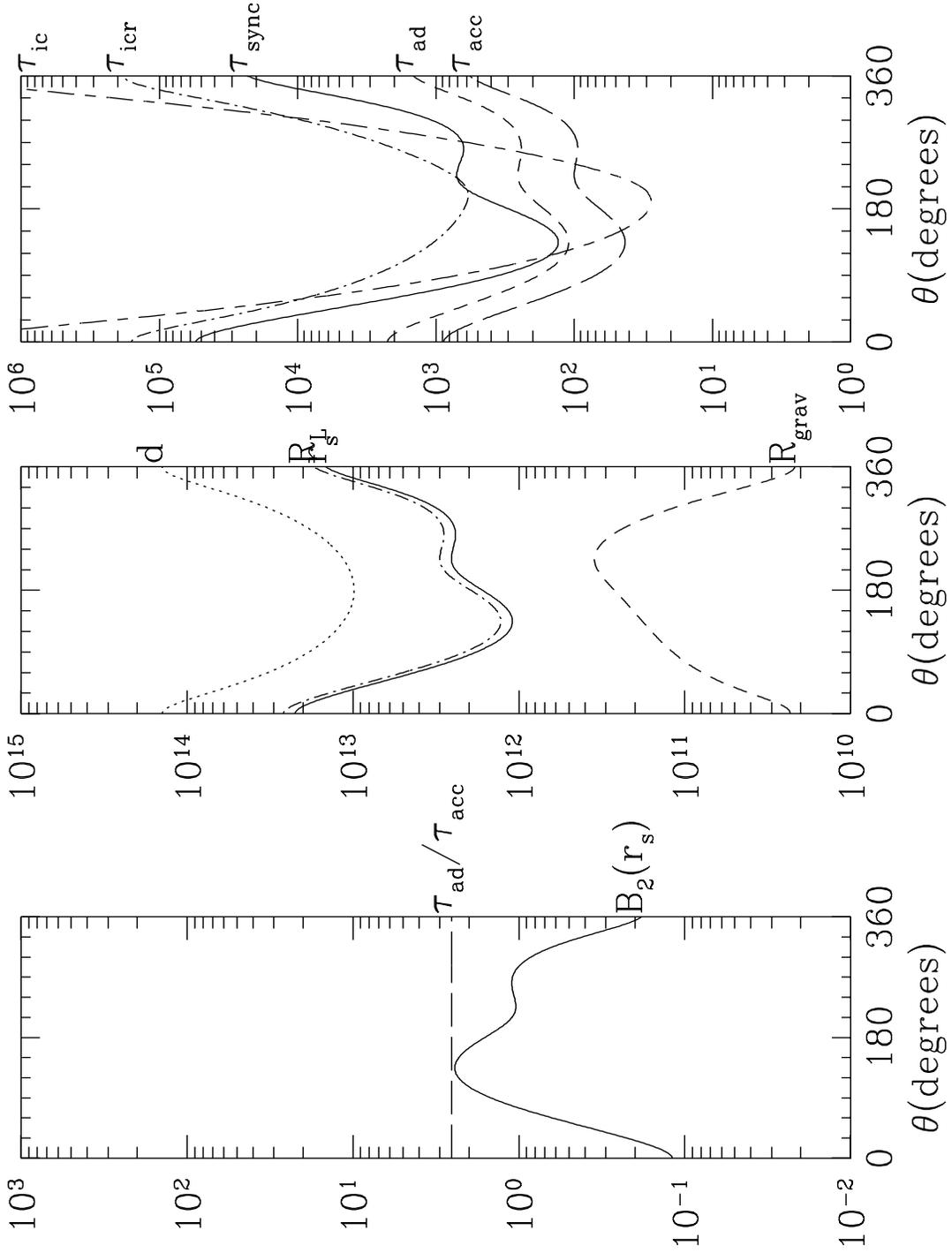

Fig. 8

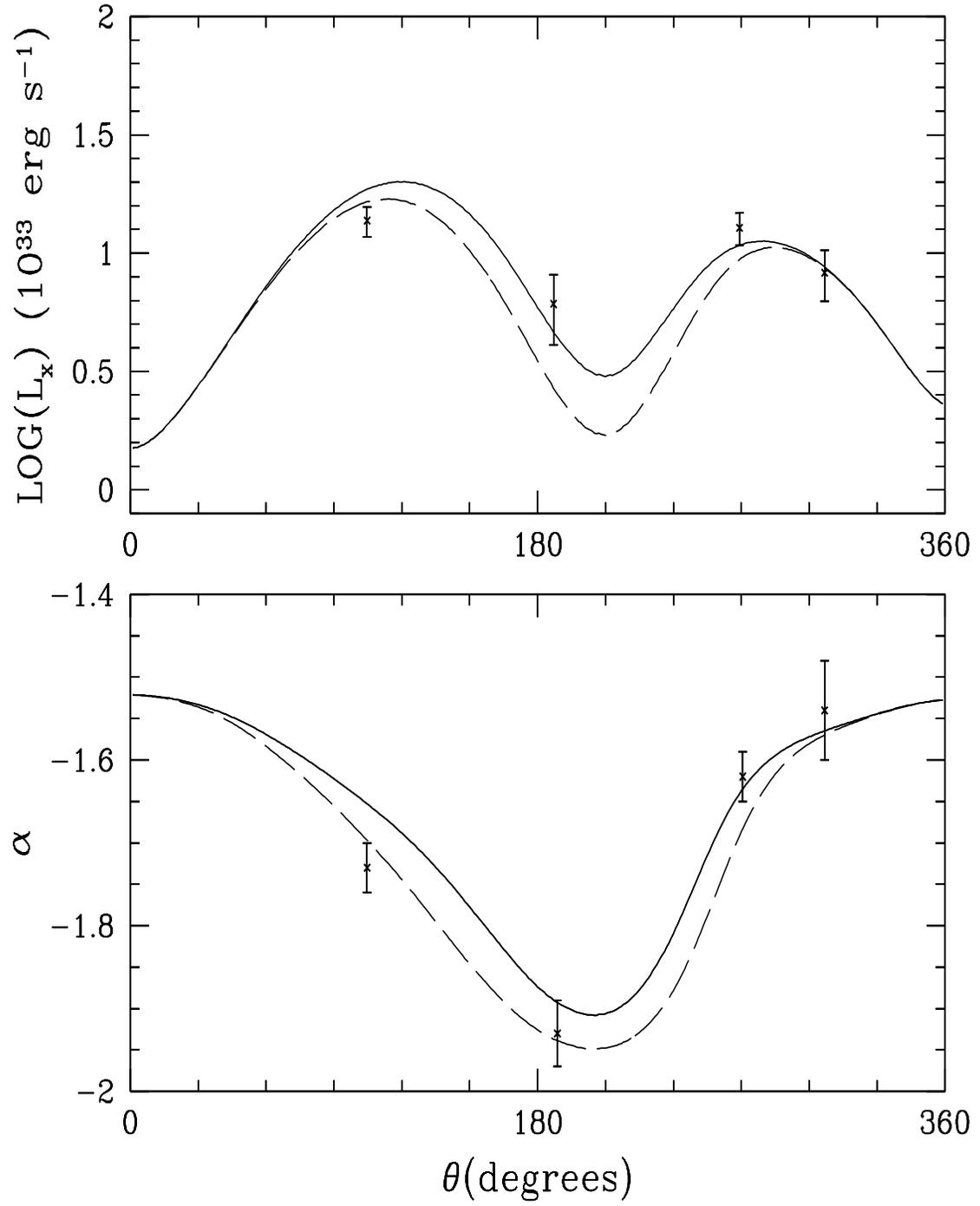

Fig. 9

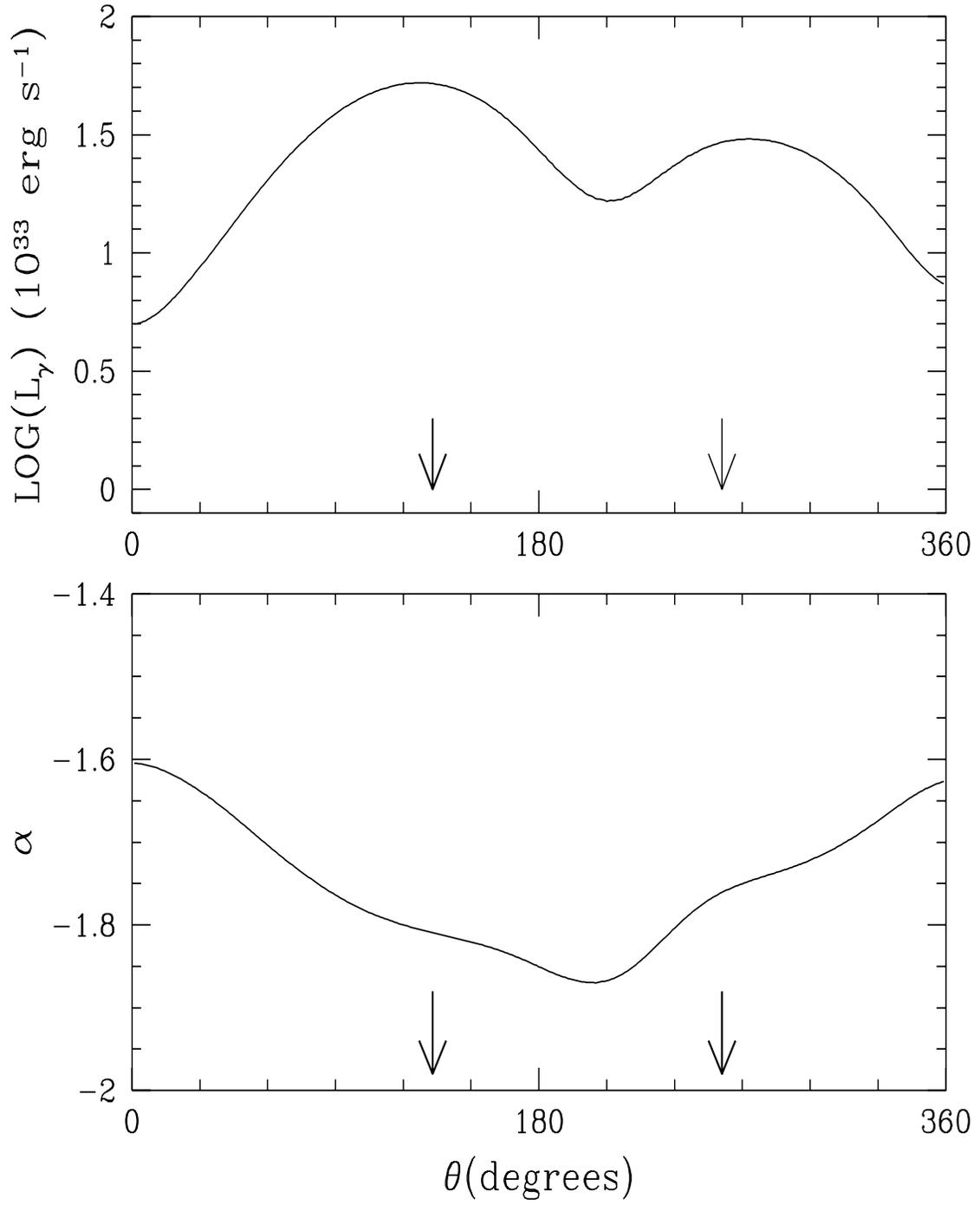

Fig. 10

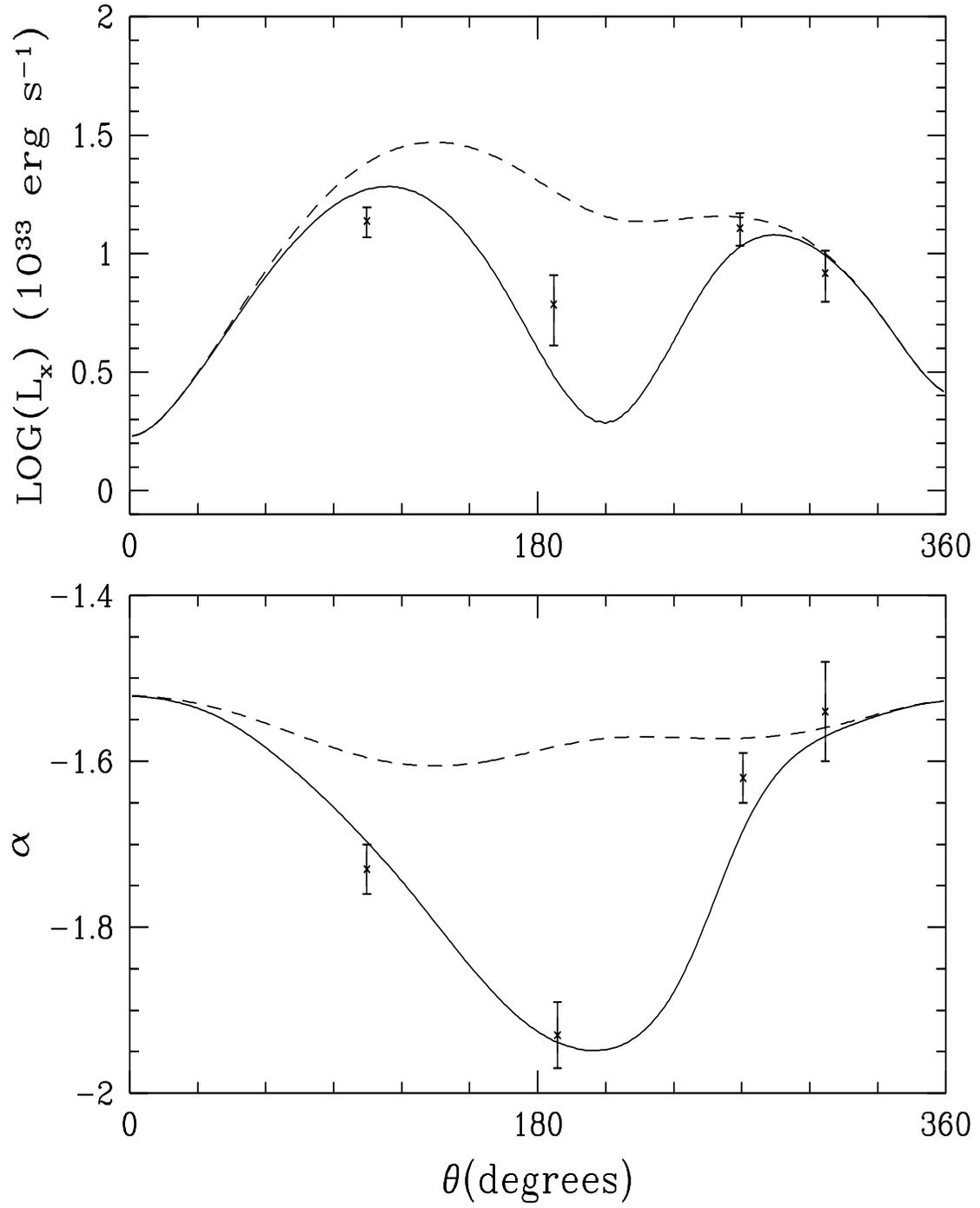

Fig. 11

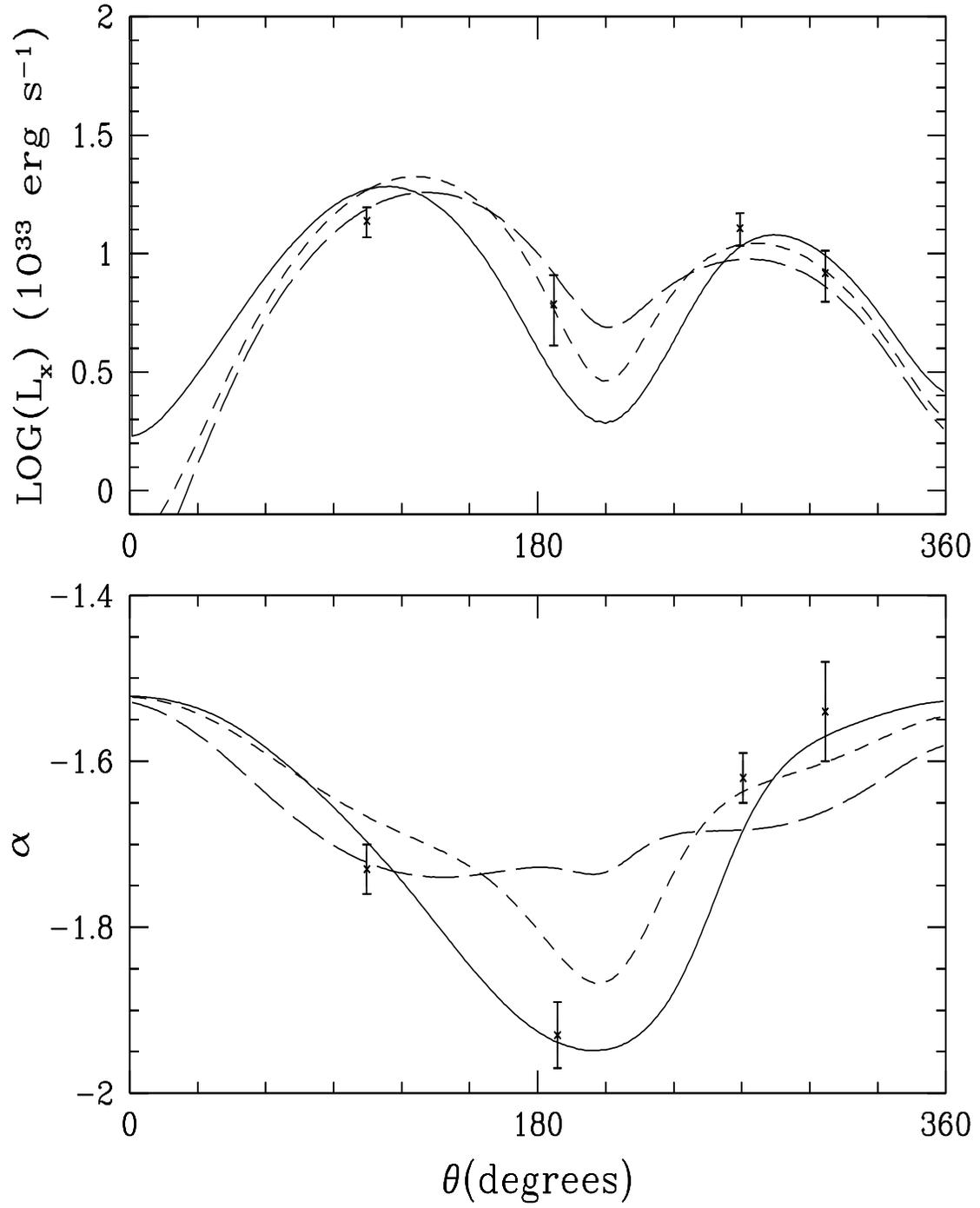

Fig. 12

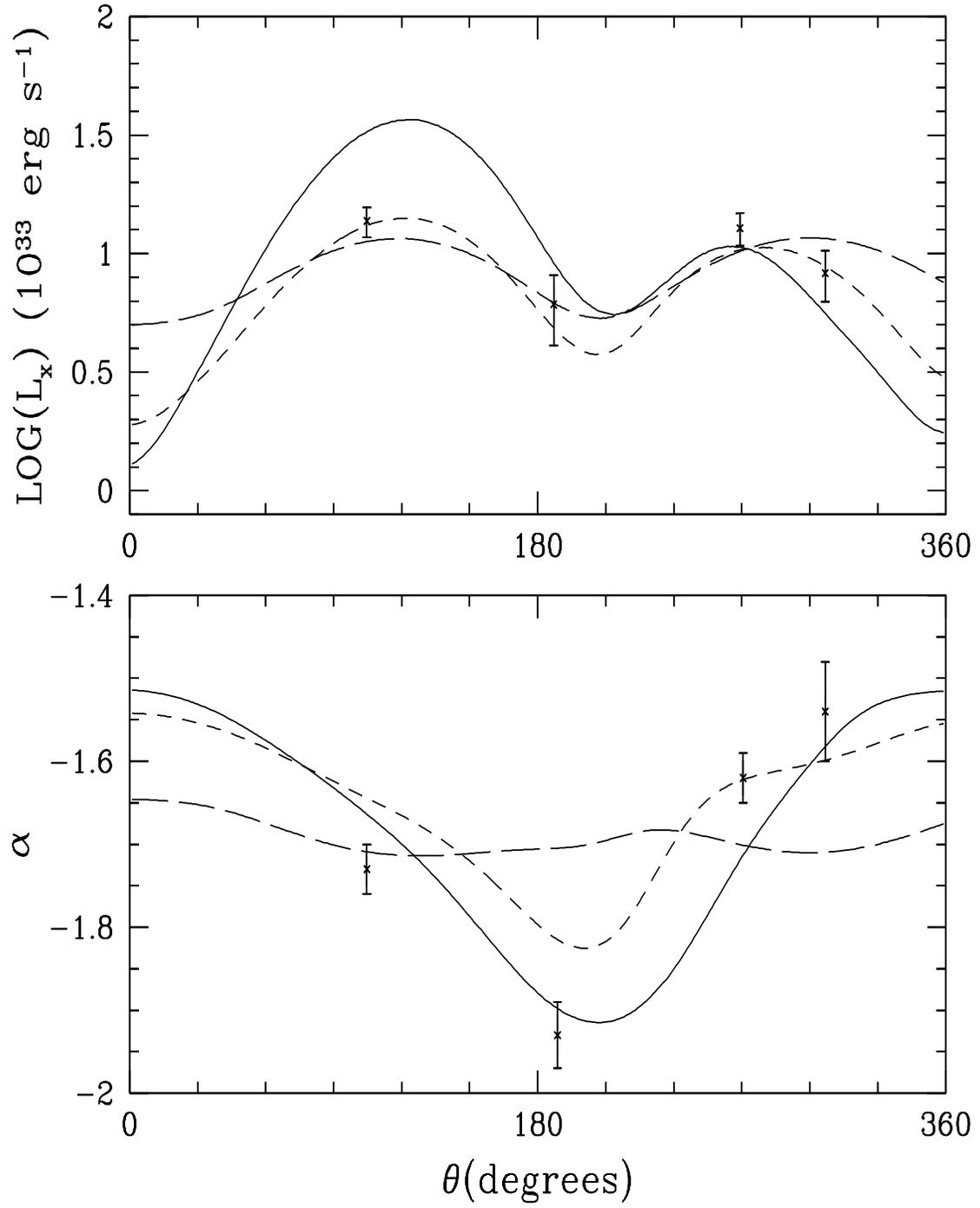

Fig. 13

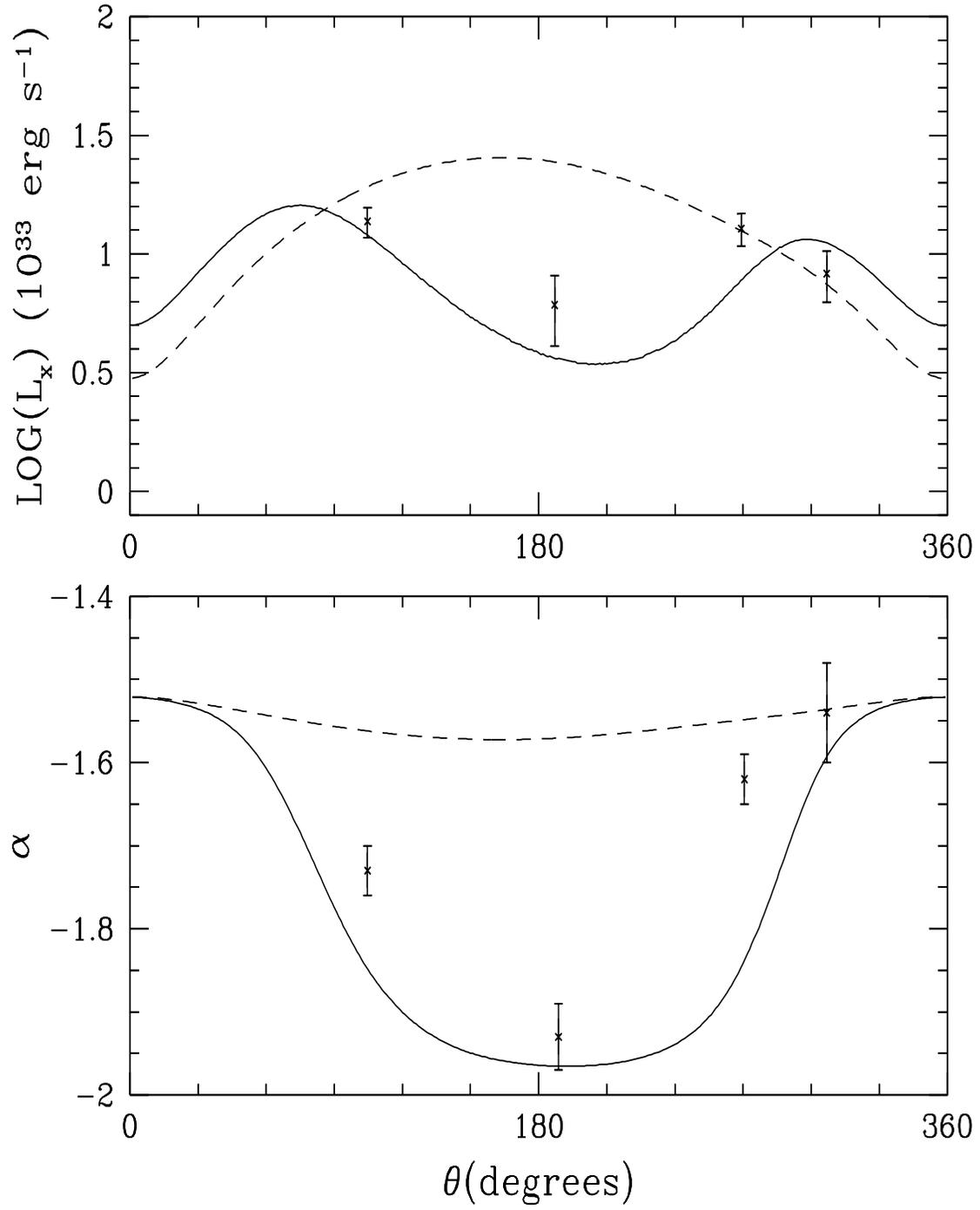

Fig. 14

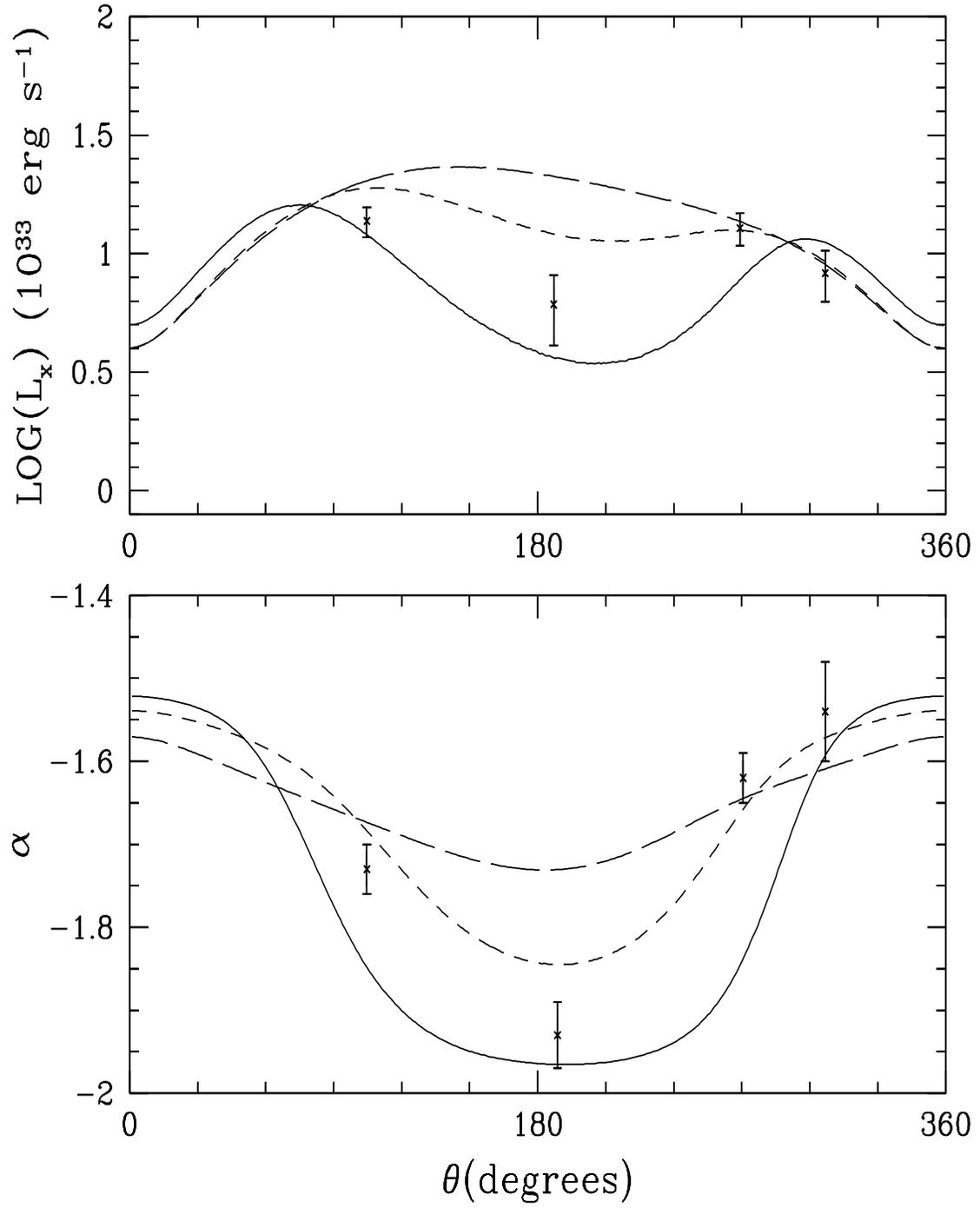

Fig. 15

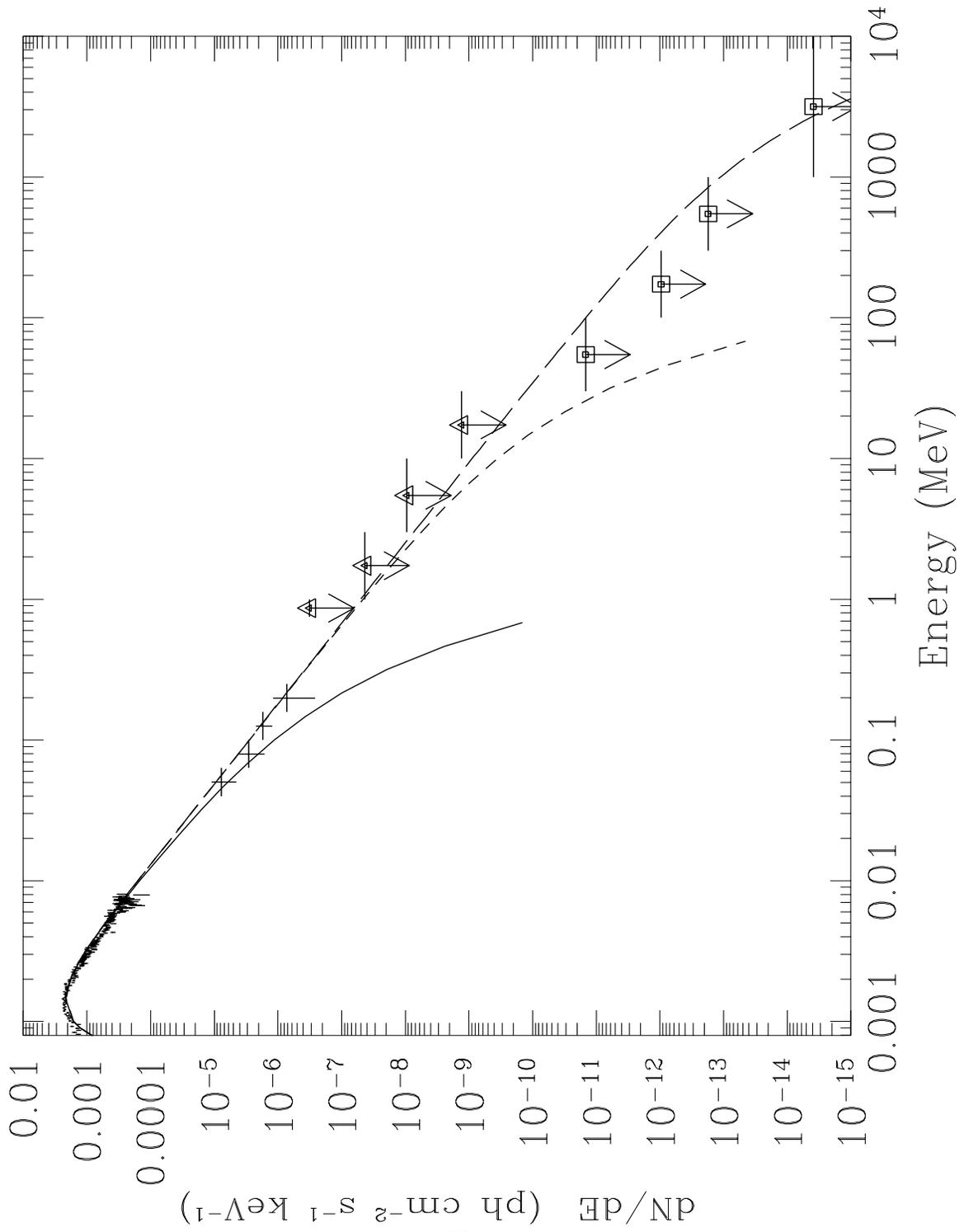

Fig. 16